# Task-dependent modulation of the visual sensory thalamus assists visual-speech recognition

**Begoña Díaz[a,b,c*], Helen Blank[b,d], and Katharina von Kriegstein[b,e]**

[a]*Center for Brain and Cognition, Pompeu Fabra University, Barcelona, 08018, Spain*

[b]*Max Planck Institute for Human Cognitive and Brain Sciences, Leipzig, 04103, Germany*

[c]*Department of Basic Sciences, Medical and Health Sciences Faculty, International University of Catalonia, Sant Cugat del Vallès, 08915, Spain*

[d]*University Medical Center Hamburg-Eppendorf, 20246 Hamburg, Germany*

[e]*Faculty of Psychology, Technische Universität Dresden, 01187 Dresden, Germany*

**Correspondence should be addressed to:**
Begoña Díaz
Postal address: Center for Brain and Cognition, Dept. Technology, C. Ramon Trias Fargas 25-27, Barcelona 08005, Spain
Telephone: +34 93 542 2629
Fax: +34 93 542 2517
Email: begona.diaz@upf.edu

# Highlights

►Recognizing visual speech, compared to face identity, increased LGN responses

►LGN modulation to speech was higher than for control tasks with non-speech stimuli

►LGN response to speech correlated positively with speechreading accuracy

►LGN modulation to speech was independent of eye-movement and task difficulty

**The cerebral cortex modulates early sensory processing via feed-back connections to sensory pathway nuclei. The functions of this top-down modulation for human behavior are poorly understood. Here, we show that top-down modulation of the visual sensory thalamus (the lateral geniculate body, LGN) is involved in visual-speech recognition. In two independent functional magnetic resonance imaging (fMRI) studies, LGN response increased when participants processed fast-varying features of articulatory movements required for visual-speech recognition, as compared to temporally more stable features required for face identification with the same stimulus material. The LGN response during the visual-speech task correlated positively with the visual-speech recognition scores across participants. In addition, the task-dependent modulation was present for speech movements and did not occur for control conditions involving non-speech biological movements. In face-to-face communication, visual speech recognition is used to enhance or even enable understanding what is said. Speech recognition is commonly explained in frameworks focusing on cerebral cortex areas. Our findings suggest that task-dependent modulation at subcortical sensory stages has an important role for communication: Together with similar findings in the auditory modality the findings imply that task-dependent modulation of the sensory thalami is a general mechanism to optimize speech recognition.**

# 1. Introduction

Sensory thalami receive massive feed-back connections from the cerebral cortex (Jones, 1985). Many studies in animals have shown that these corticothalamic connections have the power to fine-tune or even change the receptive field properties of sensory thalamic neurons (Andolina et al., 2007; Krupa et al., 1999; Lee et al., 2008; Murphy and Sillito, 1987; Sillito et al., 1994, 1993; Temereanca and Simons, 2004; Zhang et al., 1997). These findings have challenged the classical view of the sensory thalamus as a passive relay station (Camarillo et al., 2012). They opened up a debate regarding thalamic function and the behavioral relevance of top-down corticothalamic modulation (for reviews see Briggs and Usrey, 2008; Cudeiro and Sillito, 2006; Ghazanfar and Nicolelis, 2001; Ghodrati et al., 2017; Guillery and Sherman, 2002; Makinson and Huguenard, 2015; Saalmann and Kastner, 2011; Suga and Ma, 2003).

In humans, the contribution of top-down modulation of the sensory thalamus to perception and cognition is still poorly understood (Saalmann and Kastner, 2011). In the auditory modality, top-down modulation of the auditory thalamus, the medial geniculate body (MGB), is relevant for auditory speech recognition (Díaz et al., 2012; von Kriegstein et al., 2008). MGB responses increased when participants recognized fast, spectro-temporal changes in speech (i.e., speech sounds), as compared to recognizing other changes occurring at a slower time-scale in the same stimulus. The employment of the same stimuli for the two recognition tasks ensured that the MGB response increase in the speech task was driven by task requirements and not by differences in stimulus properties. In the following we will call such modulation, 'task-dependent modulation'. The amplitude of the task-dependent modulation of the MGB correlated positively with auditory speech-recognition skills (von Kriegstein et al., 2008). In the visual modality, several studies have shown task-dependent modulation of the visual thalamus, the lateral geniculate body (LGN), by attention to visual stimuli, such as flickering checkerboards or moving dots (Ling et al., 2015; O'Connor et al., 2002; Schneider, 2011; Schneider and Kastner, 2009). For example, the LGN response increased when participants attended to a peripheral flickering checkerboard as compared to when they did not attend it (O'Connor et al., 2002). LGN responses also increased when participants attended to moving white dots in contrast to stationary colored dots in a dot display (Schneider, 2011). However, whether task-dependent modulation of the LGN is relevant for human communicative functions is unknown.

Speech recognition in face-to-face communication relies on auditory- and visual-speech signals (Arnold and Hill, 2001; Navarra and Soto-Faraco, 2007; Ross et al., 2007; Sumby and Pollack, 1954). Visual-speech recognition is the ability to recognize speech based solely on the visible and fast-varying articulatory movements of the speech articulators such as the lips and the tip of the tongue. The perception of the visual-speech signal enhances the understanding of what is said by up to 45% (Arnold and Hill, 2001; Navarra and Soto-Faraco, 2007; Ross et al., 2007; Sumby and Pollack, 1954), and is particularly important in situations with high background noise (MacLeod and Summerfield, 1987) or for populations with hearing impairments (Bernstein et al., 2000; Giraud et al., 2001; Rouger et al., 2007). Yet, we also use information from visual-speech when the auditory signal is clear (Arnold and Hill, 2001; Mcgurk and Macdonald, 1976). The co-occurrence of auditory and visual information in speech opens up the question whether auditory and visual systems require similar processing mechanisms for the analysis of the fast-varying speech signals. The aim of the present study was therefore to test whether the LGN has similar response properties for visual-speech recognition as previously found for the MGB in auditory-speech recognition (Díaz et al., 2012; von Kriegstein et al., 2008). If that were the case, it would indicate that there is a unifying mechanism of task-dependent modulation of the sensory thalamus in the auditory and visual modalities: in the LGN for visual-speech recognition and in the MGB for auditory-speech recognition.

In two fMRI experiments, we measured the LGN blood oxygen level dependent (BOLD) response while participants performed a visual-speech recognition task and a face-identity task with the same visual stimulus material (i.e., muted videos of several speakers). The visual-speech recognition task required processing the fast-varying, spatio-temporal visual features of articulatory movements. The face-identity task required recognizing relatively constant visual features. The two task conditions in each experiment differed only in the specific visual feature that participants had to recognize (i.e. visual-speech or face-identity), all other aspects of the tasks where kept the same.

We tested two key hypotheses. First, we hypothesized a task-dependent modulation of the LGN with a greater BOLD response for visual-speech recognition, as compared to face-identity recognition. Second, we hypothesized a positive correlation of the task-dependent LGN modulation with visual-speech recognition task performance. Such a correlation would provide a first indication that the task-dependent LGN modulation is relevant for visual-speech recognition. The two hypotheses were derived from the analogous response characteristics in the MGB for auditory speech found in previous studies: (i) a greater BOLD response of the MGB when processing fast-varying, spectrotemporal features in auditory-speech as compared to more slowly varying voice-identity features (Díaz et al., 2012; von Kriegstein et al., 2008), and (ii) a positive correlation of the task-dependent MGB modulation with auditory speech recognition performance (von Kriegstein et al., 2008). Besides testing the two key hypotheses, we also explored whether the task-dependent modulation of the LGN was stronger for visual-speech in contrast to non-speech biological movements. To do this, we included two control conditions with non-speech biological movement stimuli in Experiment 1. In the control conditions, participants performed a task that required tracking fast movements of a thumb pressing cell phone keys in contrast to recognizing the identity of the cell phone. Thus, the control conditions involved recognizing fast (thumb movement on cell phone keys) or more stable (identity) spatio-temporal changes in a similar fashion as the visual-speech conditions. A stronger task-dependent modulation for visual speech in contrast to non-speech biological movement stimuli would be a first indication that the task-dependent modulation of the LGN is particularly relevant for the processing of movement features that are present in visual speech.

## 2. Methods

The Ethics Committee of the Medical Faculty, University of Leipzig, Germany, approved the procedures and all participants gave their written informed consent.

### *2.1. Participants, stimuli, and fMRI task design: Experiment 1*

2.1.1. Participants

Twenty-one healthy volunteers participated in Experiment 1 (native German speakers, 10 female, mean age 26 years, age range 23–34 years; all right handed according to the Edinburgh questionnaire, Oldfield, 1971). None of the participants had a history of auditory, neurological, or psychiatric disorders and they all had normal or corrected-to-normal vision. Three participants were excluded from the analysis: one because of difficulties with acquiring the field-map during fMRI, one because he did not follow the task instructions, and one due to intermittent technical problems with the response box. Therefore, the analyses were based on 18 participants (9 females; mean age 27 years).

2.1.2. Stimuli

Stimuli consisted of muted videos and of auditory-only files. Stimuli were created by recording three native German male speakers (22, 23, and 25 years old) and three cell phones. Audiovisual videos were taken of the speakers' faces and of a hand operating the cell phones. The key sounds of each cell phone had an idiosyncratic timbre and frequency (i.e., 205, 405, 470 Hz, respectively) implemented by the manufacturer of the cell phone (Supplementary Fig. 1). All the speaker videos started and ended with the speaker's mouth closed. Speech samples of each speaker included the same 12 two-syllable words (Example: "Dichter", English: "poet"). For each cell phone there were 12 videos of different sequences of two to five key presses per sequence. For the experiment, we created muted videos (mean duration 1.77 s ±0.23) and auditory-only stimuli (mean duration 0.95 s ±0.15) from the 12 speaker videos and, likewise, muted videos (mean duration 1.73 s ±0.25) and auditory-only stimuli (mean duration 1.76 s ±0.27) from the 12 cell phone videos.

Videos were recorded with a digital video camera (Canon, Legria HF S10 HD-Camcorder). High quality auditory stimuli were simultaneously recorded with a condenser microphone (Neumann TLM 50, pre-amplifier LAKE PEOPLE Mic-AmpF-35, soundcard PowerMac G5, 44.1 kHz sampling rate, and 16 bit resolution) and the software Sound Studio 3 (felt tip inc, USA). All recordings were made in a sound-attenuated chamber (IAC-I200 series, Winchester, UK) under constant luminance conditions.

All videos were processed and cut in Final Cut Pro (version 6, HD, Apple Inc., USA). They were converted to mpeg format and presented at a size of 727 × 545 pixels. In the MRI scanner, the image subtended a visual angle of 11° x 8° at a viewing distance of 98 cm. The auditory stimuli were post-processed using Matlab (version 7.7, The MathWorks, Inc., MA, USA) to adjust overall sound level. The audio files of all speakers and cell phones were adjusted to a root mean square (RMS) of 0.083.

In the following we will call all stimuli derived from the speaker recordings, person stimuli and from the cell phone recordings, cell phone stimuli.

2.1.3. FMRI experimental design

The experiment was a 2x2x2 factorial design (Fig. 1A,B) with the factors stimulus type (person stimuli/cell phone stimuli), tasks (movement task/identity task), and audiovisual (AV) congruency (AV same/AV different). In the person-stimulus conditions (Fig. 1A), each trial consisted of a visual-only video of a speaker saying a word followed by an auditory-only word said by one of the speakers. In the cell phone-stimulus condition (Fig. 1B) each trial consisted of a visual-only video of a hand pressing the keys of a cell phone followed by an auditory-only presentation of key-tones of one of the cell phones. The auditory and visual stimulus could be the same or different in content and/or identity. This means that in the person stimulus condition, the visual and auditory stimulus of each trial could represent the same or a different word and that the word could be spoken by the same or a different speaker. In the cell phone stimulus condition, the visual and auditory stimulus of each trial could represent the same or a different number of key-presses and these could be performed on the same or a different cell phone identity.

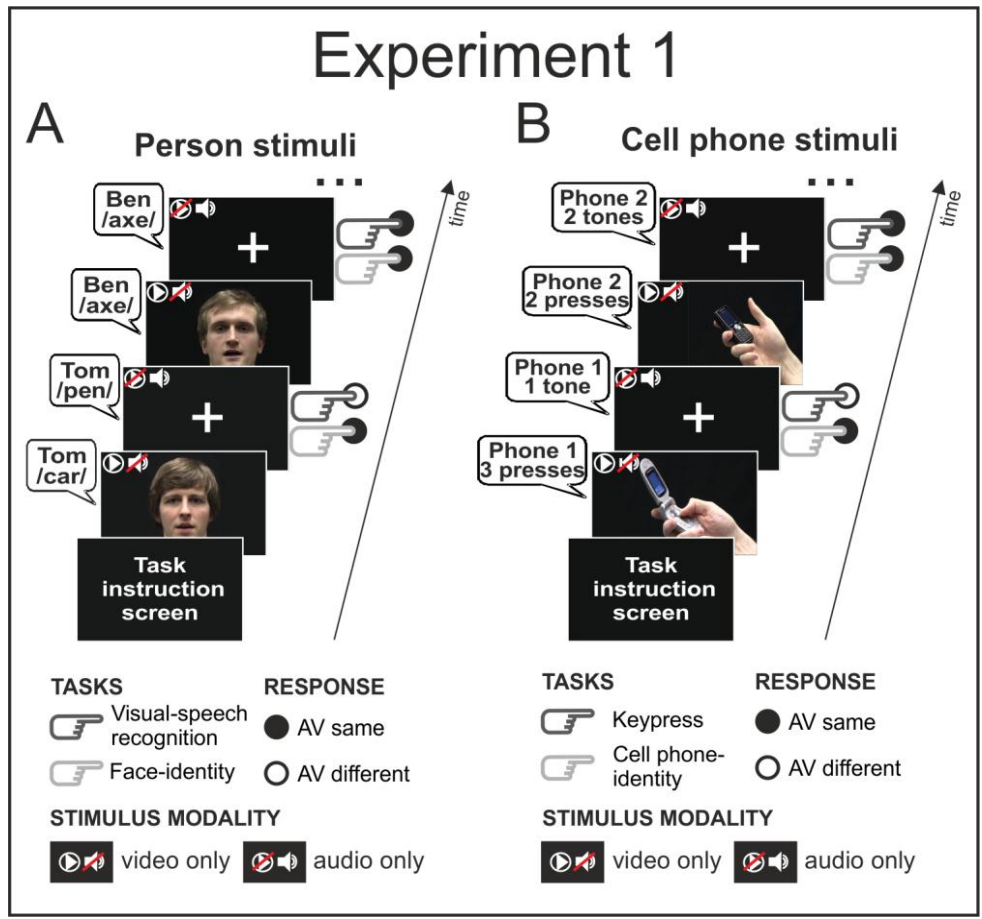

**Fig. 1. Experiment 1: Experimental Design.**
The experiment contained the factors task and stimulus condition. (A) Person stimulus tasks: One trial consisted of a muted video of one speaker (represented here by a face image) followed by the presentation of a visual fixation cross together with an auditory word. Balloons show the name of the speaker and the word. Participants performed two tasks: In the visual-speech recognition task, participants indicated whether the words in the video and audio stream were the same or different. In the face-identity task, participants indicated whether the speaker in the visual and audio stream were the same or different. Stimuli were identical for the two task conditions. Correct responses are represented by the hands color-coded for each task and color-coded circle for each type of response. Stimuli were organized in blocks of the same task condition. Before each block, participants saw a screen with task instructions. (B) Cell phone stimulus tasks: One trial consisted of a muted video of a thumb pressing cell phone keys followed by the presentation of a fixation cross together with auditory keypad tones. Balloons show the identity of the cell phones and the number of key presses/tones. Participants performed two tasks: For the keypress task, participants indicated whether the number of key presses shown in the video were the same as the number of keypad tones heard in the auditory stimulus. This task required tracking the movements of the thumb on the keypad. In the cell phone-identity task, participants indicated whether the same cell phone-identity was present in the video and audio streams. Stimuli were identical for the two task conditions. Correct responses are represented by the hands color-coded for each task and color-coded circle for each type of response. Stimuli were organized in blocks of the same task condition. Before each block, participants saw a screen with task instructions. AV, audio-visual.

Participants performed two types of tasks on the person and cell phone stimuli – a "movement" and an "identity task". In the following we will call the "movement task" in the person-stimulus condition "visual-speech recognition task" and in the cell phone-stimulus condition "keypress task". The "identity task" will be called "face-identity" and "cell phone-identity task", respectively.

In the visual-speech recognition task (Fig. 1A), participants were requested to indicate, via button press, whether the auditory-only presented word was the same as in the preceding visual-only (i.e., muted) word in the video or not. This task required visual–speech recognition during the visual-only stimulus presentation. In the keypress task (Fig. 1B), participants indicated, via button press, whether the number of auditory-only keypad tones heard was the same or different as the number of pressed cell phone keys in the muted video previously presented. In the identity task during person-stimulus conditions (face-identity task, Fig. 1A), participants indicated, via button press, whether or not the auditory-only voice and the preceding visual-only face belonged to the same person. This task required recognizing the identity of the face in the video. For the identity task during cell phone-stimulus conditions (cell phone-identity task, Fig. 1B), participants indicated whether the auditory-only key-tones and the preceding visual-only cell phone belonged to the same cell phone or not. The identity associations were learned in a training session prior to the fMRI experiment (for details about the training see the Supplementary Methods).

In total, the experiment contained 12 different speech stimuli and 12 different key-tone sequences. Each stimulus was repeated three times per task condition. The order of presentation of the word and key-tone sequences was randomized.

Within each trial, the visual-only and auditory-only stimuli were presented sequentially for 1.6 s on average each. Between the visual-only and auditory-only stimulus a fixation cross was presented (average duration 1.6 s; range 1.2 to 2.2 s). To indicate the response phase, a blue frame appeared on the screen during the auditory-only stimuli. The response phase lasted 2.3 s. Between trials, a fixation cross was presented (average 1.6 s; range 1.2 to 2.2 s). A third of the trials were null events in which a fixation cross was presented for 1.6 s. The null events were randomly presented within the experiment (Friston et al., 1999). Trials were grouped into 24 blocks of 12 trials each. There were 6 blocks per task condition. Participants performed one of the tasks in each block (i.e., visual-speech recognition task, keypress task, face-identity task, or cell phone-identity task). The grouping of trials into blocks was done to minimize the time spent instructing the participants on which task to perform and to avoid frequent task switching. Blocks were presented in pseudo-random order, not allowing neighboring blocks of the same task condition. All blocks were preceded by a short task instruction. The written words "Wort" (Eng. "word") and "Anzahl der Tastentöne" (Eng. "number of key press") indicated the visual-speech recognition and keypress tasks respectively. The written words "Person" (Eng. "person") and "Handy" (Eng. "cell phone") indicated the face- and cell phone-identity tasks respectively. Each instruction was presented for 2.7 s. The whole experiment consisted of two 16.8 min runs. Participants were allowed to rest for approximately two minutes between runs. Before the experiment, participants were briefly familiarized with all tasks outside of the MRI-scanner.

Beside the factors task (visual-speech recognition/keypress and face-identity/cell phone-identity), stimulus (person/cell phone), and AV congruency (same/different), the whole experiment included a fourth factor: modality of the first stimulus of a trial (auditory-only first/visual-only first). The "visual-only first" condition is described above. The setup for the "auditory-only first" condition was exactly the same as the "visual-only first" condition, with the difference that the first stimulus was auditory-only and the second stimulus was the visual-only video. This condition was part of a different research question and the experimental procedures and results for cortical areas are described in detail elsewhere (Blank et al., 2011; Blank and von Kriegstein, 2013). We only used the cell phone stimuli of these conditions to functionally localize the LGN independently of the conditions of interest (see below section *2.4.3.5. Definition of regions of interest (ROIs)*).

## *2.2. Participants, stimuli and fMRI task design: Experiment 2*

Experiment 2 served to replicate the findings of Experiment 1 using different stimuli and a different task design, as well as to control for potential eye-movement differences between the visual-speech recognition and face-identity task conditions (Lal and Friedlander, 1989; Sylvester et al., 2005). We used visual stimuli (i.e., vowel-consonant-vowel syllables) and a 1-back task. We used this design in the visual modality to parallel as much as possible the design of our previous studies in the auditory modality (in which we found task-dependent modulation of the MGB, Díaz et al., 2012; von Kriegstein et al., 2008).

2.2.1. Participants

Sixteen healthy volunteers participated in Experiment 2 (native German speakers, two female, mean age 27 years, age range 22–32 years; all right-handed according to the Edinburgh questionnaire, Oldfield, 1971). None of the participants had a history of auditory, neurological, or psychiatric disorders and they had normal or corrected-to-normal vision. None of them had participated in Experiment 1.

2.2.2. Stimuli

Stimuli consisted of muted videos of three native German female speakers (22, 26, and 26 years old) and a native German male speaker (22 years old). In all videos the speakers' face were shown while articulating speech and they started and ended with the speakers' mouth closed. The three female speakers were recorded while saying 54 vowel-consonant-vowel syllables. The syllables were all possible combinations of the vowels /a/, /e/, /u/, and the consonants /p/, /t/, /n/, /f/, /s/, /r/. Each of the speech sounds corresponds to a different and discriminable visual articulatory unit, commonly called visemes, in German (Aschenberner and Weiss, 2005). The video recordings of the syllables were used for the main experiment. The male speaker was recorded while saying forty 5-word sentences that were semantically neutral (e.g., ''Der Junge trägt einen Koffer'', English: "The boy carries a suitcase") and syntactically similar (i.e., subject–verb–object). The video recordings of the sentences were used for functionally localizing the LGN (see below section *2.2.3. fMRI experimental task*).

All the speakers were recorded with a digital video camera (Canon, Legria HF S10 HD-Camcorder) under constant luminance conditions. All videos were processed and cut in Final Cut Pro (version 6, HD, Apple Inc., USA). Videos of the female speakers were on average 1.89 s (± 0.19 s) long and those of the male speaker were on average 3.11 s (± 0.25 s) long. Videos were converted to mpeg format and presented at a size of 727 × 545 pixels. In the MRI-scanner, the image subtended a visual angle of 11° x 8° at a viewing distance of 98 cm.

2.2.3. FMRI experimental task

The study consisted of two parts: the main experiment and a functional localizer to identify the location of the LGN.

During the main experiment participants performed two tasks: a visual-speech recognition task and a face-identity task. The main experiment consisted of blocks of muted videos of three female speakers saying vowel-consonant-vowel syllables (Fig. 2). Blocks started with a task instruction screen (2 s) followed by a sequence of eight videos. Between the videos, a fixation cross was presented for 0.3 s at approximately the same position of the speakers' mouth. Each block lasted on average 20 s. During each block, participants performed one of the two experimental tasks according to the written instruction at the beginning of the block: "Silbe" (Eng. "syllable") for the visual-speech recognition task and "Person" (Eng. "person") for the face-identity task. For the visual-speech recognition task blocks, participants responded via button press whenever the syllable in the current video was different from the one in the previous video. For the face-identity task blocks, participants responded via button press whenever the speaker of the current video was different from the one in the previous video. Videos were randomly presented within the block with the constraints that (i) each syllable/speaker occurred at least twice, and (ii) changes between two consecutive syllables/speakers occurred between three and four times within a block. The same blocks were presented for both tasks.

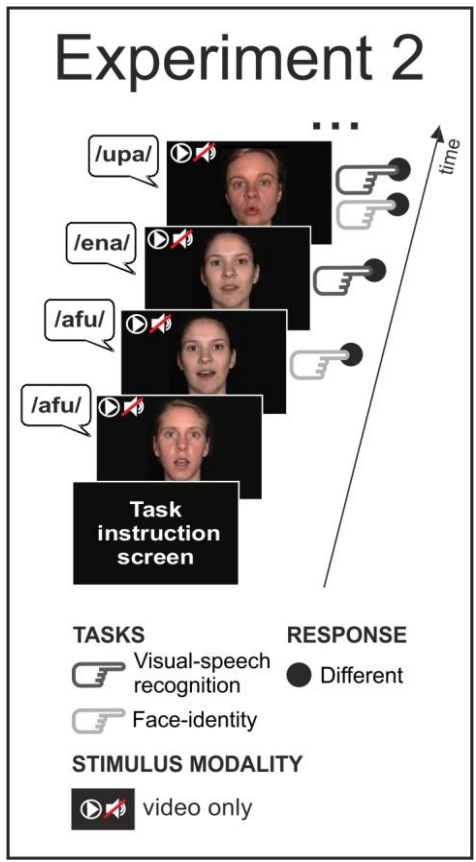

**Fig. 2. Experiment 2: Experimental Design.**
The experiment included muted videos of speakers articulating syllables and had two task conditions. In the visual-speech recognition task, participants indicated when the syllable spoken by the speaker in the muted video was different from the one in the preceding video. In the face-identity task, participants indicated when the speaker in the muted video was different from the one in the preceding video. Balloons show the syllable articulated. Correct responses are represented by the hands color-coded for each task and by a black circle for the type of response. Stimuli were organized in blocks of the same task condition. Before each block, participants saw a screen with task instructions. Stimuli were identical for the two tasks.

Participants were asked to fixate on the speakers' mouth and the fixation cross to minimize potential differences in participants' eye movements between the two tasks. In addition, a reminder to fixate on the mouth was presented at the beginning of each block below the task instruction ("Bitte auf den Mund schauen", Eng. "please, look at the mouth"). We monitored participants' eye movements via an eye tracking system (ASL Eye-Trac 6, Applied Science Laboratories, Bedford, USA).

The experiment consisted of two runs, each run had 36 blocks: 12 blocks for each task (i.e., visual-speech recognition and face-identity tasks) and 12 rest blocks, in which participants only had to look at a fixation cross for 20 s. The different types of blocks were presented in random order. Participants were allowed to rest for ca. five minutes between the two runs. The runs lasted approximately 11 minutes each. Before the experiment, participants were familiarized with the tasks outside of the MRI scanner by performing five blocks for each experimental task.

In the functional localizer for the LGN, participants were presented with 15 blocks of six randomly selected muted videos of the male speaker pronouncing a sentence. Each block of muted videos lasted on average 18 s. A fixation cross was presented for 0.4 s between the videos approximately where the speaker's mouth was located. Participants were asked to press a button at the end of each sentence to ensure that they watched the videos. The instruction was done via a task instruction displaying the written word "Satzende" (Eng. "end of the sentence") for 2 s immediately before the beginning of each block. In addition, the design included 15 rest blocks of 18 s each in which participants looked at a fixation cross. All blocks were presented in random order. The duration of the functional localizer was approximately 10 minutes.

## 2.3. Image acquisition: Experiments 1 and 2

Functional images and structural T1-weighted images were acquired on a 3 T Siemens Tim Trio MR scanner with a 12-channel head coil (Siemens Healthcare, Erlangen, Germany).

For the functional MRI, a gradient-echo EPI (echo planar imaging) sequence was used (TE 30 ms, flip angle 90 degrees, TR 2.79 s, 42 slices, whole brain coverage, acquisition bandwidth 116 kHz). The voxel size was 3 mm$^3$ (2 mm slice thickness, 1 mm interslice gap, in-plane resolution 3 mm × 3 mm), a resolution sensitive to capture signals from small brain structures such as the LGN (Kastner et al., 2004; Mullen et al., 2010; Noesselt et al., 2010; O'Connor et al., 2002; Sylvester et al., 2005; Wunderlich et al., 2005). For Experiment 1, 403 volumes were acquired. The volumes covered the presentation of all experimental conditions including also those that we used for functionally localizing the LGN. For experiment 2, 524 volumes were acquired for the main experiment (two runs) and 228 for the functional localizer (one run). Geometric distortions were characterized by a B0 field-map scan. The field-map scan consisted of gradient-echo readout (24 echoes, inter-echo time 0.95 ms) with standard 2D phase encoding. The B0 field was obtained by a linear fit to the unwrapped phases of all odd echoes.

The structural images were acquired with a T1-weighted magnetization-prepared rapid gradient echo sequence (3D MP-RAGE) with selective water excitation and linear phase encoding. Magnetization preparation consisted of a non-selective inversion pulse. The imaging parameters were TI = 650 ms, TR = 1300 ms, TE = 3.93 ms, alpha = 10°, spatial resolution of 1 mm$^3$, two averages. To avoid aliasing, oversampling was performed in the read direction (head–foot).

## *2.4. Data analysis: Experiments 1 and 2*

2.4.1. Quantification of visual movement

We estimated the amount of movement present in the person and cell phone stimuli of Experiment 1 by tracking the position in the x- and y-axis of the upper and lower lips for the person stimuli and the tip of the thumb for the cell phone stimuli on a frame-by-frame level. Two coders that were blind to the aim of the study performed the tracking by means of the software Tracker (http://www.cabrillo.edu/~dbrown/tracker/) developed by the Open Source Physics project(Christian et al., 2011). We run statistical analyses (SPSS 18.0, SPSS Inc., Chicago, IL, USA) on four measures: the average displacement in the x- and y-axis between consecutive frames, and the average velocity and acceleration parameters. All these parameters were computed automatically by the software Tracker for each video frame. For the person stimuli, the values for the upper and lower lips were added to obtain a global estimation of the movement parameters.

2.4.2. Behavioral data

*2.4.2.1. Task performance*

We computed the percentage of correct responses for each task and participant and transformed them to rationalized arcsine units (RAU) to alleviate ceiling effects (Studebaker G.A., 1985). Statistical tests were performed by means of SPSS 18.0 (SPSS Inc., Chicago, IL, USA) to assess for potential differences in task difficulty between the movement and identity tasks. The normal distribution of the data was assessed with the Shapiro-Wilk normality test. All datasets satisfied the normality assumption and we compared task performance by means of paired sample t-test. In addition we compared participants' performance across the two experiments by means of a multivariate ANOVA with the between-subjects factors "Experiment" (1 and 2) and the within-subjects factor "Task" (visual-speech recognition and face-identity task).

*2.4.2.2. Eye tracking data*

For experiment 2, we also collected eye tracking data to test whether LGN modulation could be caused by differences in eye movements between the visual-speech recognition and face-identity recognition tasks (Lal and Friedlander, 1989; Sylvester et al., 2005). Eye data were analyzed offline with ASL-software (ASL Results Plus, Applied Science Laboratories, Bedford, USA). We used SPSS 18.0 (SPSS Inc., Chicago, IL, USA) for statistical comparisons. We run two analyses on the eye data. Firstly, we tested whether there were differences in the average number and duration of fixations between the visual-speech recognition and face-identity tasks. For one participant there were no eye data because of difficulties with the calibration of the eye tracker. Hence, the analysis included the data from 15 participants. Secondly, we analyzed whether the location of the fixations on the experimental videos differed depending on the task. For this analysis we assessed the average number and duration of fixations on the experimental stimuli for three different regions of interest: the speakers' mouth, eyes, and nose (Supplementary Fig. 2). In this second analysis we included the data from a subset of the participants, 7, for whom we obtained reliable localization of the fixations. The specific details about the eye tracking analyses are reported in the Supplementary Methods.

### 2.4.3. Functional MRI data

*2.4.3.1 Preprocessing*

MRI data were analyzed with SPM8 (v5236; Wellcome Trust Centre for Neuroimaging, UCL, London, UK, www.fil.ion.ucl.ac.uk/spm) in a Matlab environment (version 9.2.0.556344, R2017a) (The MathWorks). EPI-Scans were realigned and unwarped to correct for motion artifacts. To achieve high precision in intersubject alignment we spatially normalized the images to the MNI template by means of the Diffeomorphic Anatomical Registration Through Exponentiated Lie algebra (DARTEL) procedure (Ashburner, 2007). For this procedure, we co-registered the participants' structural images to a participant's mean of the realigned functional images. The structural images were then segmented into tissue-class images for gray matter, white matter, and cerebrospinal fluid (New Segment in SPM8). The participants' tissue-class images were used to create a mean structural template representative of all participants' brains and to calculate the deformation fields (i.e., flow fields) for each participant's native space image to the common space. Functional images were normalized to MNI space by mapping the group structural template to an MNI template in combination with the participant specific deformation fields. The normalized functional images preserved intensities of the original images (i.e., no "modulated"). Images were spatially smoothed with a Gaussian smoothing kernel of 4 mm full width at half maximum (FWHM). Smoothing is necessary to increase the signal-to-noise ratio, compensate for between-subject variability, and to normalize error distributions to permit application of Gaussian random field theory for the statistics inference (Friston et al., 2000). Additionally, we also ran analyses without spatially smoothing the data and with a smoothing kernel of 2 mm to check whether smoothing misplaced the local maxima (Mikl et al., 2008). Geometric distortions due to susceptibility gradients were corrected by an interpolation procedure based on the B0 map (the field-map).

*2.4.3.2. Design matrices*

For Experiment 1, statistical parametric maps were generated by modeling the evoked hemodynamic response of the events of interest separately for each condition as boxcar functions convolved with a synthetic hemodynamic response function using the general linear model approach (Friston et al., 2007). The events of interest were the visual stimuli of the visual-speech recognition task, face-identity task, keypress task, and cell phone-identity task for the visual-first condition. In the same model, we included the events used to localize the LGNs (see below section *2.4.3.5. Definition of regions of interest (ROIs)*). The rest of the events were not modeled. For Experiment 2, statistical parametric maps were generated by modeling the evoked hemodynamic response for the visual-speech recognition and face-identity task blocks as boxcar functions convolved with a synthetic hemodynamic response function using the general linear model approach (Friston et al., 2007).

*2.4.3.3. Categorical analyses*

The contrast of interest to test our main hypothesis, i.e., that LGN responses are modulated by the visual-speech recognition task, was "visual-speech recognition task – face-identity task". We computed this contrast for each participant at the first-level by means of a t-contrast. At the second-level, we used a one-sample t-test across the first-level contrast images of all participants. We performed two different second-level analyses on the contrast of interest. For Analysis 1, we created a design matrix with the first-level contrast images of all participants. For Analysis 2, we created a design matrix with the first-level contrast images of all participants and we entered as covariate of no interest the difference in correct responses (in RAU) between the visual-speech and face-identity tasks to control for task difficulty. For Experiment 2 only, Analysis 2 included as additional covariates of no interest the average number of eye fixations and the duration of eye fixations to control for potential eye-movement effects. The categorical analysis of Experiment 1 also allowed exploring whether the task-dependent modulation of the LGN was stronger for visual-speech in contrast to non-speech biological movements. To do that we computed the interaction between stimulus type and task by means of an F contrast in SPM "(visual-speech recognition task/person stimuli – face-identity task/person stimuli) – (keypress task/cell phone stimuli – cell phone-identity task/cell phone stimuli)". The F contrast was computed at the second-level based on the contrast images of each condition (compared against the implicit baseline) computed for each participant at the first-level.

*2.4.3.4. Correlation analyses*

We used correlation analyses to test our second hypothesis, i.e. that the task dependent modulation of the LGN correlates positively with visual-speech recognition accuracy. We created a design matrix at the second-level with the contrast "visual-speech recognition task – face-identity task" of each participant and each participants' behavioral score (in RAU) for the visual-speech recognition task as covariate of interest. We ran an additional exploratory analysis to check whether there is a positive correlation between the BOLD response for the visual-speech recognition task and the participants' visual-speech recognition score. To do that we created a design matrix at the second-level with the contrast images "visual-speech recognition task – implicit baseline" of each participant and each participant's behavioral score (in RAU) for the visual-speech recognition task as a covariate of interest. We expected no positive correlation between LGN responses in the other conditions and behavioral accuracy in these conditions and checked this by computing the correlations between the BOLD response for each of the other tasks and the respective behavioral scores.

*2.4.3.5. Definition of regions of interest (ROIs)*

We used the functional localizers to define group-based ROIs combined with probabilistic cytoarchitectonic maps of the LGNs (from the Anatomy toolbox v.2.2c for SPM, Eickhoff et al., 2005). Intersubject averaging is a reliable tool to assess LGN responses (Büchel, Turner, & Friston, 1997) and a group-based ROI approach has been used for the study of other small, subcortical structures such as the MGB (Díaz et al., 2012; Thompson et al., 2006; von Kriegstein et al., 2008). In addition, in the present study we used DARTEL to optimize intersubject registration. For Experiment 1, the LGNs were functionally mapped by contrasting the conditions with videos of cell phones from the auditory-only first conditions (not used for the experimental contrasts) against the implicit baseline at the first-level and a one-sample t-test across the first-level contrast images of all participants at the second-level. This means that the LGN localizer was independently defined from the contrast of interest used to address our hypotheses. For Experiment 2, the LGNs were localized by contrasting the blocks of muted videos of the male speaker saying sentences against the implicit baseline at the first-level and a one-sample t-test across the first-level contrast images of all participants at the second-level. These analyses were run on the 4 mm, 2 mm, and unsmoothed data. Following previous studies (Mullen, Thompson, & Hess, 2010; Schneider & Kastner, 2009; Wunderlich, Schneider, & Kastner, 2005), the LGNs were defined as all contiguous voxels responsive to the videos and located in the anatomical position of the LGN (Table 1 provides the location of the maximum statistic of the LGN clusters, which were similar to the ones reported by previous studies as showed in Supplementary Table 1). The ROI was defined by the intersection of the functional clusters with the probabilistic cytoarchitectonic maps of the visual thalamus provided by the Anatomy toolbox for SPM (version 2.2c) (Eickhoff et al., 2005) (Supplementary Fig. 3). The ROIs were created with the Marsbar Toolbox and exported to the functional image space (http://marsbar.sourceforge.net).

**Table 1. Definition of the LGN regions of interest (ROIs).**

|  | ROIs | MNI coordinates | Volume (mm$^3$) |
|---|---|---|---|
| **Experiment 1** | rLGN | 24, -27, -3 | 88 |
|  | lLGN | -21, -27, -3 | 232 |
| **Experiment 2** | rLGN | 27, -24, -3 | 176 |
|  | lLGN | -21, -27, -3 | 216 |

The MNI coordinates of the local maxima in millimeters are x, y, z. rLGN and lLGN stand for the right and left lateral geniculate nucleus, and ROI stands for region of interest.

2.5. Statistical thresholds

For the quantification of the movement in the videos, the behavioral response, and eye tracking data analyses, effects were considered significant if present at p < 0.05. For the fMRI analyses (categorical and correlation analyses), effects were considered significant if present at P < 0.05 family-wise error (FWE) corrected for multiple comparisons at the peak level within the regions of interest (i.e., right and left LGN). We computed the FWE correction for the ROIs by means of small volume correction.

## 3. Results

### *3.1. Visual-speech recognition modulates LGN responses*

To address our first hypothesis, we tested whether the LGN response was modulated by the visual-speech recognition task in contrast to the face-identity task in the person stimulus conditions of Experiment 1 (Fig. 1A). In the visual-speech recognition task, participants indicated whether a word spoken in a muted video matched a subsequently presented auditory word or not. In the face-identity task, participants indicated whether the identity of the speaker in the muted video matched the identity of the voice in a subsequently presented auditory sample (Fig. 1A). The stimuli for both task conditions were exactly the same. The experiment also included cell phone stimulus tasks (Fig. 1B).

As hypothesized, we found that the BOLD response in left and right LGNs (Fig. 3A) was significantly higher during the visual-speech recognition task, as compared to the face-identity task (Fig. 3B and 2C, Table 2 "Analysis 1, categorical/person stimuli").

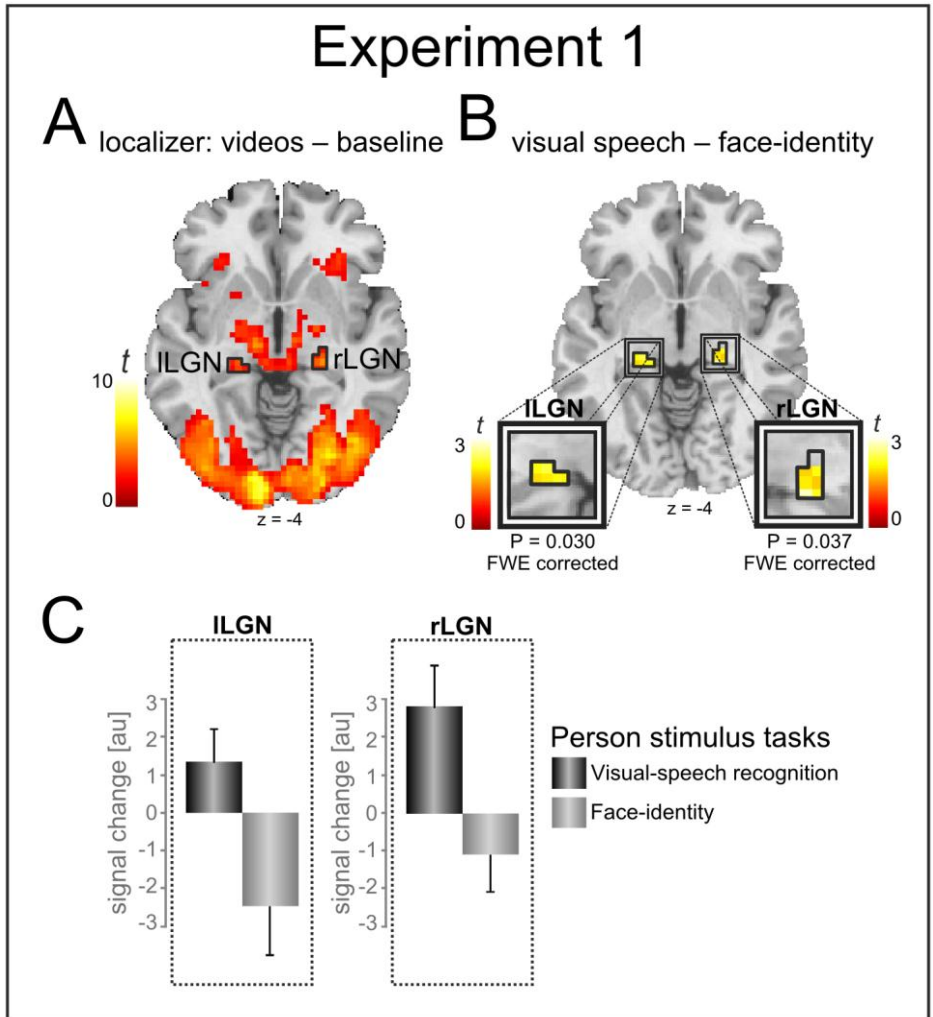

**Fig. 3. Experiment 1: fMRI results for person stimulus conditions.**
(A) We identified the LGNs based on the independent localizer conditions (shown here) and an intersection with probabilistic maps (see supplementary Figure 3). (B) Within the LGN ROIs, there were higher responses to the visual-speech recognition in contrast to the face-identity task. Significance testing was done in the statistical parametric mapping software (SPM). For visualization purposes only, we masked the results with the LGN clusters from the visual localizer (indicated by the black outline surrounding the LGN). (C) The bar plots serve to illustrate the amount of LGN response for each person stimulus task condition separately. The bars depict the parameter estimates (au: arbitrary units) at the local response maxima of the contrast between visual-speech recognition and face-identity tasks (MNI coordinates: 24, -27, -6 and -21, -27, -6). Error bars indicate SEM.

**Table 2. Experiment 1: Local fMRI response maxima and statistics.**

|  |  | Analysis 1 | | | Analysis 2 |
|---|---|---|---|---|---|
|  |  | *Categorical/ person stimuli* <br> *'visual-speech rec. – face-identity'* | *Categorical/ Interaction* <br> *Task x stimuli type* | *Categorical/ cell phone stimuli* <br> *'keypress – cell phone-identity'* | *Categorical/ person stimuli* <br> *'visual-speech rec. – face-identity'* |
| **rLGN** | coordinates | 24, -27, -6 | 24, -27, 6 |  | 24, -27, -6 |
|  | P value | 0.037, FWE | 0.025, FWE | n.s. | 0.041, FWE |
|  | Z score | 2.25 | 2.45 |  | 2.43 |
| **lLGN** | coordinates | -21, -27, -6 | -24, -27, -6 |  | -21, -27, -6 |
|  | P value | 0.030, FWE | 0.033, FWE | n.s. | 0.037, FWE |
|  | Z score | 2.56 | 2.54 |  | 2.48 |

The MNI coordinates of the local maxima in millimeters are x, y, z. Analysis 2 differs from Analysis 1 in that it additionally included a covariate of no interest with the behavioral performance difference (percentage of correct responses in RAU) between the visual-speech and face-identity task to control for differences in task difficulty. rLGN, right lateral geniculate nucleus; lLGN, left lateral geniculate nucleus; FWE, family-wise error corrected for region of interest; n.s., not significant even at a lenient statistical threshold (P < 0.05, uncorrected).

### 3.2. LGN modulation for visual-speech but not for non-speech biological (*thumb*) movements

We next explored whether the task-dependent LGN modulation is higher for the person stimuli (Fig. 1A) than for the cell-phone stimuli (Fig. 1B), which contained non-speech biological movement. The tasks in the cell-phone stimulus conditions were similar to the person stimulus tasks in that one required to focus on the movement present in the stimulus (keypress task), and the other on the identity of the cell phone (cell phone-identity task; for more details see legend of Fig. 1B). We performed an interaction analysis between task conditions and stimulus type. The interaction "(visual-speech recognition task/person stimuli – face-identity task/person stimuli) – (keypress task/cell phone stimuli – cell phone-identity task/cell phone stimuli)" was significant within the left and right LGN (Fig. 4, Table 2 "Analysis 1, categorical/interaction"). The direction of the interaction was as expected: The LGN responses were higher for the visual-speech recognition task as compared to the face-identity task (Table 2 "Analysis 1, categorical/person stimuli"). There was however no significant response for the keypress task in contrast to the cell phone-identity task (Table 2 "Analysis 1, categorical/cell phone stimuli"), even at a lenient statistical threshold (P < 0.05, uncorrected).

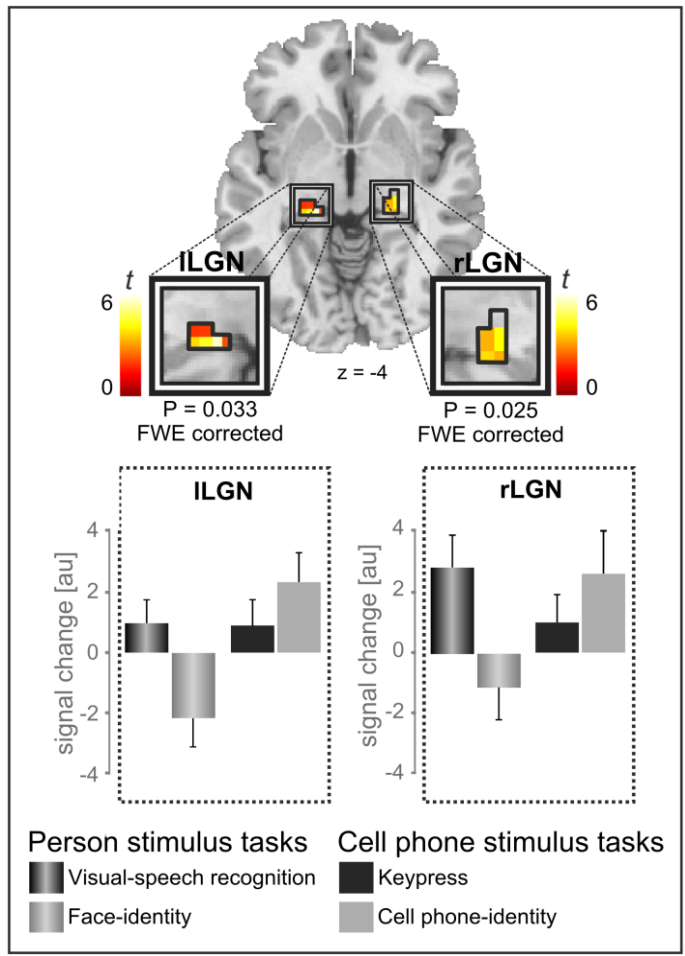

**Fig. 4. Experiment 1: fMRI results for interaction between task and stimulus conditions.**
Within the LGN ROIs, there was a significant interaction between task and stimulus conditions: "(visual-speech recognition task/person stimuli – face-identity task/person stimuli) – (keypress task/cell phone stimuli – cell phone-identity task/cell phone stimuli)". Significance testing was done in the statistical parametric mapping software (SPM). For visualization purposes only, we masked the results with the LGN clusters from the visual localizer (indicated by the black outline surrounding the LGN). The bar plots serve to illustrate the amount of LGN response for each condition separately. The bars depict the parameter estimates (au: arbitrary units) at the local response maximum of the interaction between task and stimulus conditions (MNI coordinates: 24, -27, -6 and -24, -27, -6). Error bars indicate SEM.

The entire lack of task-dependent modulation for the cell phone stimuli was somewhat unexpected, as a previous study had shown that attention to moving stimuli (white dots) in contrast to attention to static stimuli (colored dots) leads to LGN BOLD response increase (Schneider, 2011). To check whether the lack of task-dependent modulation for the cell phone stimuli in our study might be explained by less movement in these stimuli compared to the person stimuli, we analyzed the amount of movement and speed in the videos. The results showed that the cell phone stimuli contained even larger and faster movements than the person stimuli (Table 3). Thus the lack of task-dependent modulation of the LGN for the cell phone stimuli cannot be accounted for by less movement present in the cell phone stimuli in comparison to the person stimuli.

**Table 3. Experiment 1: Quantification of visual movement in person and cell phone stimuli.**

|  | **Person stimuli** | **Cell phone stimuli** | **Two-sample t-test** |
|---|---|---|---|
| **X-displacement** | 0.25 ± 0.07 pixels | 4.39 ± 0.82 pixels | $t(70) = 30.05, p < 0.001$ |
| **Y-displacement** | 1.04 ± 0.27 pixels | 9.63 ± 1.11 pixels | $t(70) = 44.92, p < 0.001$ |
| **Velocity*** | 0.86 ± 3.54 m/s | 2.85 ± 0.29 m/s | $t(70) = 3.35, p = 0.001$ |
| **Acceleration*** | 15.86 ± 68.21 m/s$^2$ | 33.59 ± 4.39 m/s$^2$ | $t(70) = 1.55, p > 0.05$ |

The table displays the average movement measures ±standard deviations for each stimulus type (person and cell phone) and statistical comparisons between the stimulus types. *Velocity and acceleration measures do not correspond to the world coordinates but to the image size in pixels. We used an arbitrary correspondence of 1 pixel = 1 centimeter.

### *3.3. Task-dependent LGN modulation is independent of task difficulty*

In Experiment 1, the visual-speech recognition task was significantly more difficult than the face-identity task (Table 4). We therefore investigated whether task difficulty differences could explain the LGN modulation. We included a behavioral measure for task difficulty (i.e., percentage correct responses in RAU for the visual-speech recognition task – face-identity task) as a covariate of no interest in the analysis. Participants showed large individual variability in their performance pattern across the tasks: The task difficulty covariate ranged from 14.81 to -43.58 RAU (mean: -12.97±14.37). When we included the task difficulty covariate the results for the task contrast stayed qualitatively the same (Table 2 "Analysis 2, categorical/person stimuli"). Furthermore the amount of task difficulty did not correlate with the amount of task-dependent LGN modulation even at a lenient statistical threshold (P < 0.01, uncorrected). In addition, for the cell phone conditions (Fig. 1B), the cell phone-identity task was significantly more difficult than the keypress task (Table 4). Despite this difference in difficulty, there was no task-dependent LGN modulation for the contrast "cell phone-identity task – keypress task" even at a lenient threshold (P < 0.01, uncorrected). In summary, it seems very unlikely that the task-dependent modulation of LGN responses can be explained by differences in difficulty between the visual-speech and face-identity task: (i) there was no correlation between task difficulty and LGN modulation for the person stimuli, (ii) there was no task-dependent LGN modulation for the cell phone conditions although there was a statistically significant differences in task performance, and (iii) the task-dependent modulation of the LGN for the person stimuli remained the same when difficulty differences were entered as covariate of no interest in the model.

**Table 4. Experiment 1: Behavioral performance.**

| Visual-speech recognition task | Face-identity task | Paired samples t-test |
|---|---|---|
| RAU: 82.79 ±13.33<br>W(18) = 0.95, p = 0.50<br>% hits: 81.60 ± 10.81 | RAU: 95.76 ±13.67<br>W(18) = 0.94, p = 0.39<br>% hits: 90.47 ± 7.87 | t(17) = 3.82, p = 0.001 |
| **Keypress task** | **Cell phone-identity task** | **Paired samples t-test** |
| RAU: 96.35 ±13.66<br>W(18) = 0.92, p = 0.16<br>% hits: 90.93±8.28 | RAU: 87.33 ±14.01<br>W(18) = 0.92, p = 0.15<br>% hits: 85.02±10.70 | t(17) = 2.1, p = 0.042 |

The table displays the percentage of correct responses in rationalized arcsine units (RAU) ±standard deviations and the normality tests (Shapiro-Wilk normality test: W statistic and p value) for each task condition. For completeness we also provide the untransformed percentage of correct responses ±standard deviations. The last column provides the statistical comparisons between the tasks.

### *3.4. LGN responses during visual-speech recognition correlated with visual-speech recognition performance*

We next tested our second hypothesis, i.e., that the amount of LGN modulation correlates positively with visual-speech recognition performance. Against our hypothesis, the correlation was not significant even at a lenient statistical threshold ($P < 0.05$, uncorrected). We therefore explored whether there was a positive correlation between LGN responses and behavioral responses for the visual-speech task. This was indeed the case: left LGN: P value = 0.041, FWE corrected, MNI coordinates = -21, -27, -6, Z = 2.43; right LGN: P value = 0.034 FWE corrected, MNI coordinates = 21, -27, -3, Z = 2.54 (Fig. 5). It is unlikely that the positive correlation between LGN responses and visual-speech recognition scores can be explained by potential compensatory responses during difficult visual tasks. The difficult tasks in this experiment were the visual-speech task and the cell phone-identity task. If the LGN response was related to compensatory mechanisms in difficult visual tasks, we would expect similar correlations between the LGN response and behavioral scores for the cell-phone identity task and the visual-speech task. This was however not the case. There was no correlation for the LGN responses in the cell-phone identity task and the behavioral scores in the cell phone identity task (left LGN: P = 0.223, FWE corrected; right LGN: P = 0.582, FWE corrected; both p<0.05 uncorrected). As expected, there were also no correlations between the LGN responses in the other two conditions (face-identity task; keypress task) with their respective behavioral scores (all p values were > 0.05, uncorrected and >0.258, FWE corrected).

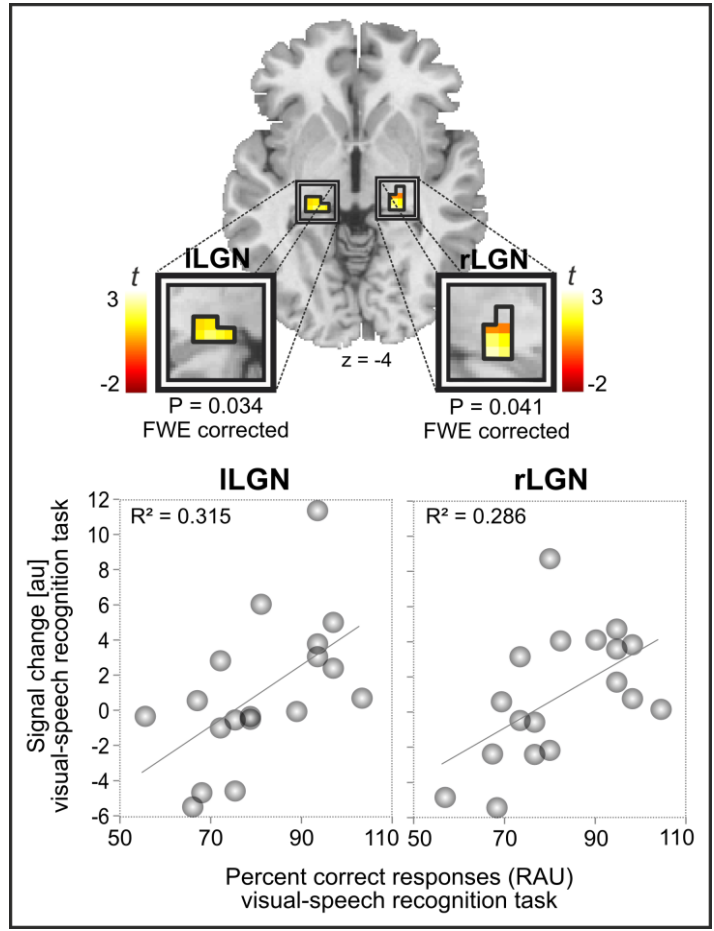

**Fig. 5. Experiment 1: Correlation analysis results.**
There was a significant positive correlation between the LGN responses during the visual-speech recognition task (au: arbitrary units) and participants' accuracy in the visual-speech recognition task (in RAU). Significance testing was done in the statistical parametric mapping software (SPM). For visualization purposes only, we masked the results with the LGN clusters from the visual localizer (indicated by the black outline surrounding the LGN). For the scatter plots we used the parameter estimates (au: arbitrary units) at the local response maximum of the correlation (MNI coordinates: -21, -27, -6 and 21, -27, -3). The scatter plots were produced for visualization purposes and SPSS was used only to calculate $R^2$.

### 3.5. Task-dependent LGN modulation is independent of eye-movements

We performed Experiment 2 to replicate the findings of the task-dependent LGN modulation with a different design and to control for eye movements. The design included sequences of muted videos from several persons (Fig. 2). Participants performed a visual-speech recognition and a face-identity one-back task on the same stimulus material. In the visual-speech recognition task, participants indicated when the presented video had different speech content than the previous one. In the face-identity task, participants indicated when the person in the video was of a different identity from the person in the previous video. To control for eye movements, participants were asked to fixate on the same point of the screen (i.e., the mouth of the speakers) across the two task conditions and eye movements were monitored with an MRI-compatible eye tracker system.

The results of Experiment 2 were similar to those of Experiment 1. Analysis 1 showed that the right LGN (Fig. 6A) had a significantly higher BOLD response to the visual-speech recognition task as compared to the face-identity task (Fig. 6B and 6C, Table 5 "Analysis 1").

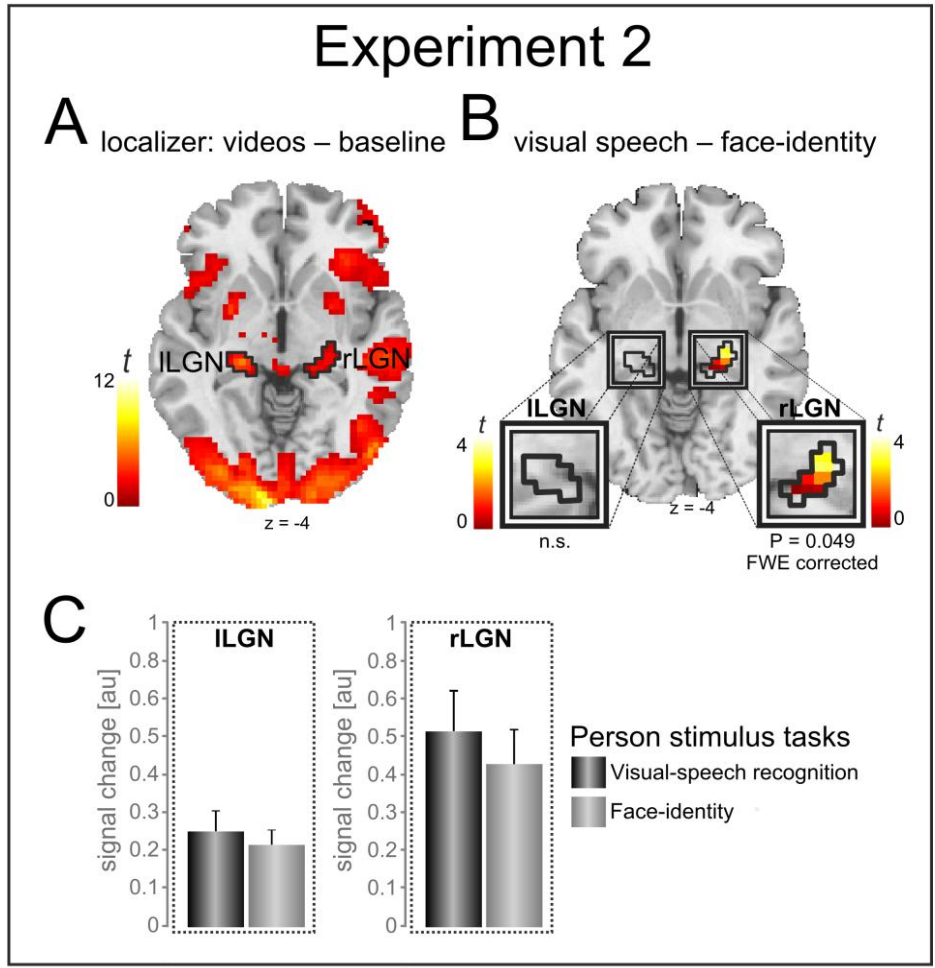

**Fig. 6. Experiment 2: fMRI results.**
(A) We identified the LGNs based on the independent localizer conditions (shown here) and an intersection with probabilistic maps (see supplementary Figure 3). (B) There was a significantly higher response to the visual-speech recognition in contrast to the face-identity task in the right LGN. Significance testing was done in the statistical parametric mapping software (SPM). For visualization purposes only, we masked the results with the LGN clusters from the visual localizer (indicated by the black outline surrounding the LGN). (C) The bar plots serve to illustrate the amount of LGN response for each condition separately. The bars display the parameter estimates (au: arbitrary units) at the local response maximum for the contrast between visual-speech recognition and face-identity tasks (MNI coordinates: 24, -27, -3 and -27, -27, -3). Error bars indicate SEM.

**Table 5. Experiment 2: Local fMRI response maxima and statistics.**

|  |  | Analysis 1 | Analysis 2 |
|---|---|---|---|
|  |  | *Categorical*<br>*'visual-speech rec. –*<br>*face-identity'* | *Categorical*<br>*'visual-speech rec. –*<br>*face-identity'* |
| **rLGN** | *coordinates* | 24, -27, -3 | 24, -27, -3 |
|  | *P value* | 0.049, FWE | 0.046, FWE |
|  | *Z score* | 2.16 | 2.46 |
| **lLGN** | *P value* | n.s. | n.s. |

The MNI coordinates of the local maxima in millimeters are x, y, z. Analysis 2 differs from Analysis 1 in that it additionally included covariates of no interest with (i) the behavioral performance difference between the visual-speech and face-identity task to control for differences in task difficulty, (ii) the average number of eye fixations, and (ii) the duration of eye fixations. rLGN, right lateral geniculate nucleus; lLGN, left lateral geniculate nucleus; FWE, family-wise error corrected for region of interest; unc., uncorrected; n.s., not significant.

LGN responses can be modulated by eye-movement and fixations (Lal and Friedlander, 1989; Sylvester et al., 2005). In the eye-tracking data we therefore first checked whether there was a difference in the average number and duration of fixations during the visual-speech recognition task and the face-identity task. The tasks did not significantly differ in the number of fixations (paired t-test: t(14) = 1.85, p = 0.085; visual-speech: 13.84 ±6.63; face-identity: 15.48 ±7.50) and the duration of the fixations (paired t-test: t(14) < 1, p = 0.57; visual-speech: 2.09 s±1.14; face-identity: 2.04 s±1.09). We checked whether these statistically non-significant differences had nevertheless an influence on the LGN responses. To do this, we calculated the difference in the average number and duration of fixations between the two tasks (visual-speech recognition – face-identity task) for each participant and entered them as covariates of no interest in Analysis 2 at the second-level. Analysis 2 also included the task difficulty as regressor of no interest (i.e., the behavioral difference between the visual-speech recognition and face-identity task performance for each participant; Table 6). The task difficulty covariate ranged from -1.94 to -24.03 RAU (mean: -12.67±7.19). The fMRI results for Analysis 2 stayed qualitatively the same, that is, there was a significantly higher BOLD response in the right LGN for the visual-speech recognition, as compared to the face-identity task (Table 5 "Analysis 2, categorical"). In addition, the average number of fixations and the duration of fixations did not correlate with the task-dependent modulation of the right LGN, even at a lenient statistical threshold (P < 0.05, uncorrected).

**Table 6. Experiment 2: Behavioral performance.**

| Visual-speech recognition task | Face-identity task | Paired samples t-test |
|---|---|---|
| RAU: 91.73 ±8.12<br>W(16) = 0.91, p = 0.12<br>% hits: 89.55 ± 5.56 | RAU: 104.41 ±6.55<br>W(16) = 0.95, p = 0.59<br>% hits: 96.61 ± 2.66 | t(15) = 7.04, p < 0.001 |

The table displays the percentage of correct responses in rationalized arcsine units (RAU) ±standard deviations and the normality tests (Shapiro-Wilk normality test: W statistic and p value) for each task condition. For the sake of completeness we also provided the untransformed percentage of correct responses ±standard deviations. The last column provides the statistical comparisons between the tasks.

In a next step, we checked the fixation location in seven participants for which these data were available (Supplementary Methods). To do that, we defined 3 ROIs on the face of the speaker corresponding to the eyes, nose, and mouth (Supplementary Fig. 2). There were no significant differences between the visual-speech recognition and face-identity tasks for the fixation location (main effect of experimental task: F(1,6) < 1; experimental task x ROI: F(1,6) < 1). There was a main effect of ROI, indicating that participants did not look equally to all the regions of the speakers' face (main effect of ROI: F(1,6) = 23.51, p < 0.001). As instructed, participants fixated more on the mouth region (mean fixations per block: 7.68 ±1.65) as compared to the nose (paired t-test: t(6) = 3.74, mean fixations per block: 3.01 ±3.15, p = 0.01) and eye regions (paired t-test: t(6)

= 10.66, p < 001; mean fixations per block: 0.75 ±0.62,). Overall, fixations were slightly longer (+0.1 s) for the visual-speech task, as compared to the face-identity task, but the difference was not significant (main effect of tasks: $F(1,6) = 4.48$, $p = 0.079$; visual-speech recognition: 0.077 s ±0.38; face-identity: 0.66 s ±0.10). We cannot exclude that the other participants for whom fixation location data was not available behaved differently. We therefore checked (for Analysis 2) whether the task-dependent modulation of the LGN was present in the seven participants for whom we can be sure that they fixated on the mouth in both tasks. In this analysis, we found a trend towards significantly higher BOLD response to the visual-speech recognition task as compared to the face-identity task in the right LGN ($P = 0.063$, FWE corrected; $Z = 2.08$; MNI coordinates: 27, -27, -3). Altogether, the results suggested that the task-dependent LGN modulation is independent of the amount and duration of fixations and not caused by different fixation locations between the tasks.

In Experiment 2, no significant correlation was found between participants' visual-speech recognition accuracy and the LGN responses . We speculate that the lack of correlation between BOLD response and behavior in Experiment 2 was due to the lower inter-individual variability in performance for the visual-speech recognition task in Experiment 2 in contrast to Experiment 1 (Tables 3 and 5).

Experiment 2 seemed to yield weaker results than Experiment 1, as only the right LGN showed significant task-dependent modulation in Experiment 2 while the task-dependent modulation in Experiment 1 was significant in the right and the left LGN. To check whether this descriptively weaker task-dependent modulation of the LGN in Experiment 2 could be explained by task difficulty levels, we tested for differences in task difficulty for the visual-speech recognition and face-identity tasks between the two experiments This was done with a multivariate ANOVA with the between-subjects factors "Experiment" (1 and 2) and within-subjects factor "Task" (visual-speech recognition and face-identity task). Overall, participants performed better in Experiment 1 than Experiment 2 (main effect of experiment: $F(1,64) = 10.69$, $p < 0.01$) and they also performed better in the face-identity task compared to the visual-speech task (main effect of task: $F(1,64) = 22.74$, $p < 0.001$). However, there was no interaction between experiments and tasks (experiment x task: $F(1,64) < 1$), indicating that the difficulty difference between visual-speech recognition and face-identity task was comparable in Experiment 1 and 2. This suggested that the descriptively weaker results in Experiment 2 were not related to a smaller task difficulty difference in Experiment 2 in comparison to Experiment 1.

### *3.6. Task-dependent modulation is specific to signals within the LGN ROIs*

We ran further analyses for both experiments to ensure that the task-dependent modulation was due to response changes in the LGNs and not due to signal spread from neighboring regions. The LGN is a relatively small structure (i.e., volume ca. 115-121 mm$^3$, Andrews et al., 1997). The pulvinar is directly adjacent to the LGN and several of its nuclei have been reported as being task-dependent modulated (Arcaro et al., 2015; Kastner et al., 2004; Schneider, 2011). Particularly, the ventral pulvinar 1 and 2 are close to the LGN (Arcaro et al., 2015). A neuroimaging study in humans has shown that ventral pulvinar 1 and 2 are located more medial, posterior, and superior than the LGN and their centers are in total approximately 10-14 mm away from that of the LGN (Arcaro et al., 2015). We checked whether the LGN modulation in our study was due to signal of neighboring regions displaced as consequence of data smoothing (Mikl et al., 2008). To investigate the effects of smoothing on the location of the maximum statistics we reanalyzed the data with less and no smoothing applied to the functional MR-images. For some effects, coordinates of the local maxima varied slightly (in total 3-6 mm) depending on the smoothing applied (Supplementary Table 2). Yet, in no case the displacement of the maximum statistic was large enough to correspond to the ventral pulvinar nuclei (>10 mm) nor in the direction of their location (the displacements were never towards more medial, posterior, and superior locations as expected for the ventral pulvinar nuclei, Arcaro et al., 2015). The location of the maximum statistics for the different smoothing strongly suggested that the task modulation by visual speech and its correlation with accuracy in visual-speech recognition originate from the LGN.

## 4. Discussion

In two fMRI experiments we compared the responses of the visual sensory thalamus, the LGN, when participants performed a task which required processing fast spatio-temporal changes of faces for visual-speech recognition or a task which required processing more stable spatio-temporal features of faces used for face identification. The tasks were performed on exactly the same stimulus material in order to avoid any differences driven by the sensory input. Furthermore, in both experiments, all the task conditions, experimental and control, had very similar structure; hence, the multiple stages of the task that are not exclusively dedicated to speech processing were controlled for as much as possible. The design and hypotheses of our study were motivated by two previous fMRI studies on auditory-speech recognition (Díaz et al., 2012; von Kriegstein et al., 2008). These studies showed that the BOLD responses in the auditory thalamus, the MGB, increased when participants processed fast-varying, spectro-temporal changes of auditory-speech, as compared to other features of the same auditory-speech signal, and the MGB modulation correlated positively with speech recognition skills (von Kriegstein et al., 2008). In the present study, we found that our findings in the visual modality by-and-large parallel those of the auditory modality: The two independent experiments confirmed our first hypothesis that LGN response is enhanced when recognizing visual-speech, as compared to processing more stable features of the same visual stimuli. The replication of this task-dependent LGN modulation in two independent studies with different experimental designs, task structure, video material, and participants speaks for the reproducibility and the reliability of the findings. Our second hypothesis was partly confirmed. There was a correlation of the visual-speech task with the LGN response during visual speech recognition, and we take this finding as a first indication that the amount of LGN response might be relevant for performance accuracy in visual-speech recognition. However, we did not find the initially hypothesized positive correlation of the task-dependent LGN modulation during visual-speech recognition with the performance of the visual-speech recognition task. We had expected such a correlation as we had found a correlation between task-dependent MGB modulation and performance in auditory-speech recognition in two independent experiments previously (von Kriegstein et al., 2008). It is an open question why we did not find a similar correlation in the present experiment for the LGN and visual-speech recognition. It could be based on an inherent difference between functioning of the LGN for visual-speech recognition and the MGB for auditory-speech recognition or on methodological differences between the studies

    The hypothesis that the LGN has a similar function for visual-speech recognition to that of the MGB in auditory-speech recognition was not self-evident given the many differences between the two thalamic structures. The LGN is mainly composed of six layers, the two ventral layers contain magnocellular neurons and the four dorsal layers

contain parvocellular neurons (Malpeli and Baker, 1975). The MGB is subdivided into three parts—the ventral, medial, and dorsal divisions—each containing different cell types (Winner, 1984). The LGN has a retinotopic representation (Chen et al., 1999; Malpeli and Baker, 1975), whereas the MGB is partially tonotopically organized (Aitkin and Webster, 1972; Moerel et al., 2015; Rouiller et al., 1989). The LGN receives visual input directly from the complex neuronal system of the retina, and the MGB receives the auditory input from a hierarchy of brainstem nuclei. Thus, the LGN and MGB not only differ in the type of sensory input they receive, visual or auditory, but also in their histological structure and organization. Albeit these differences, our findings imply that the LGN has similar task-dependent modulation during visual-speech recognition as the MGB has for auditory-speech recognition (Díaz et al., 2012; von Kriegstein et al., 2008).

The task-dependent LGN modulation by visual speech was independent from variables such as task difficulty or eye gaze behavior, which could potentially modulate LGN responses (O'Connor et al., 2002; Sylvester et al., 2005). For instance, a more difficult task could lead to more attention allocation to the stimulus (Ling et al., 2015; O'Connor et al., 2002; Schneider, 2011; Schneider and Kastner, 2009). However, our control analyses showed that the task-dependent LGN modulation was unrelated to task difficulty or to compensatory mechanisms during difficult visual tasks in better performers. Task difficulty was also unrelated to the descriptively weaker results in Experiment 2 than Experiment 1. We also controlled for potential effects of eye gaze differences between the visual-speech recognition and the face-identity tasks as previous fMRI studies showed that repetitive and large (i.e., 35 degrees) eye movements lead to a decrease of LGN BOLD activity as compared to fixation (Lal and Friedlander, 1989; Sylvester et al., 2005). The eye-tracking results and control analyses of Experiment 2 showed that the task-dependent modulation of the LGN was independent from eye gaze behavior. Experiment 2 also controlled for potential effects of the over-representation of central vision in the LGN (Azzopardi and Cowey, 1996; Schneider et al., 2004) by asking participants to fixate on the same part of the visual input (i.e., the speakers' mouth) during the two tasks of interest, the visual-speech recognition and face-identity tasks. Participants processed the same foveal and peripheral visual information during the two tasks but a stronger right LGN response was present for the visual-speech recognition tasks. Thus, the present LGN modulation cannot be fully explained by the magnification of the fovea relative to the periphery in the LGN.

Modulation of thalamic responses according to relevant features of the sensory input has been repeatedly reported by electrophysiological recording studies in animals across sensory modalities (Andolina et al., 2007; Cudeiro and Sillito, 1996; Krupa et al., 1999; Murphy and Sillito, 1987; Sillito et al., 1993; Zhang et al., 1997). However, the computational mechanism as well as the behavioral relevance of this modulation is controversially discussed (Camarillo et al., 2012; Saalmann and Kastner, 2011; von Kriegstein et al., 2008). Sillito and colleagues (1994) showed feature-linked synchronized firing of LGN neurons with non-overlapping receptive fields aligned to the orientation of moving visual stimuli. No synchronization was present when the cortex was removed or the neurons were simultaneously stimulated by flashing or drifting squares. Based on these cortical-feedback dependent response properties of LGN neurons to object movement, Sillito and colleagues (1994) proposed that the cerebral cortex generates "hypotheses" about the incoming movement trajectory against which the sensory input is then tested. Such a view is reminiscent of computational hierarchical models and predictive coding accounts of brain function that posit that internal models of the environment are dynamically exploited by the brain to predict sensory input (Friston, 2005; Kiebel et al., 2008). Predictive coding accounts have also been extended to explain speech processing (Blank and Davis, 2016; van Wassenhove et al., 2005; von Kriegstein et al., 2008), where it is assumed that the speech recognition system reaches fast and robust processing by generating predictions about the incoming input. The optimal encoding of relatively fast-temporal dynamics in the visual and auditory modalities is important for speech recognition (Campbell et al., 1997; Shannon et al., 1995). The trajectories of these fast changes in the sensory input are highly predictable. For example, because of co-articulation, the features of a speech sound/movement already predict the features of the following speech sound/movement (Jesse and Massaro, 2010; Warren and Marslen-Wilson, 1987). We speculate that such predictions could be implemented by task-dependent modulation of the LGN to fine-tune visual processing according to expected visual changes in a dynamic, iterative process. Similar mechanisms could also be present in the auditory modality (Díaz et al., 2012; von Kriegstein et al., 2008). The LGN and the MGB are tuned to high frequencies, between 10 and 20 Hz (Giraud et al., 2000; Hicks et al., 1983), whereas visual and auditory cortices are tuned to frequencies between 4 and 8 Hz (Foster et al., 1985; Giraud et al., 2000). The very high temporal resolution of the sensory thalamus combined with the fast transfer of information between thalamus and cortex trough the corticothalamic loop (around 52 ms/loop) (Briggs and Usrey, 2007) renders the sensory thalamus an ideal structure for optimized processing of fast-varying components of auditory- and visual-speech.

In a predictive coding view, task-dependent modulation of the sensory thalamus would be particularly relevant in situations where the stimulus is predictable. Such a view might reconcile the findings of the present as well as previous studies: There is task-dependent modulation of sensory thalami for auditory (Díaz et al., 2012; von Kriegstein et al., 2008) and visual speech (present study) as well as (predictable) dot movements (Schneider, 2011). In contrast we found no task-dependent modulation for tracking the relatively unpredictable movement of a thumb pressing the keys of a cell phone. To directly test the hypothesis that task-dependent modulation of the LGN is dependent on the predictability of the stimuli is, unfortunately, not possible with the present data set. The findings, however, highlight the need to systematically study the exact conditions under which task-dependent modulation of the LGN (and also the MGB, Díaz et al., 2012; von Kriegstein et al., 2008) occurs.

The present findings fundamentally contribute to our so far very scarce knowledge (Díaz et al., 2012; Schneider, 2011; von Kriegstein et al., 2008) on the behavioral relevance of task-dependent thalamus modulation: They suggest an important role of task-dependent modulation of the visual sensory thalamus for analyzing speech—one of the most important and complex signals that humans are faced with. The present and previous findings (Díaz et al., 2012; von Kriegstein et al., 2008) challenge current neuroscience models (Friederici and Alter, 2004; Hickok and Poeppel, 2007; Poeppel et al., 2012), which explain speech perception by-and-large on the basis of cerebral cortex areas. The similar task-dependent modulation of the sensory thalamus in the auditory (Díaz et al., 2012; von Kriegstein et al., 2008) and visual modalities for speech recognition however implies that a full understanding of speech perception might need to take dynamic corticothalamic interactions into account.

**Acknowledgments:** We would like to thank S. J. Kiebel for his comments on an earlier version of the manuscript. A Max Planck Research Group Grant and an ERC-Consolidator Grant (SENSOCOM, 647051) to K.v.K. supported this work. B.D. received funding from the People Programme (Marie Curie Actions) of the European Union's Seventh Framework Programme (FP7/2007–2013, REA grant agreement No. 32867) and a Juan de la Cierva fellowship (JCI-2012-12678). The authors declare to have no competing financial interests.

# References

Aitkin, L.M., Webster, W.R., 1972. Medial geniculate body of the cat: Organization and responses to tonal stimuli of neurons in ventral division. J. Neurophysiol. 35, 365–380.

Andolina, I.M., Jones, H.E., Wang, W., Sillito, A.M., 2007. Corticothalamic feedback enhances stimulus response precision in the visual system. Proc. Natl. Acad. Sci. U. S. A. 104, 1685–1690. doi:10.1073/pnas.0609318104

Andrews, T.J., Halpern, S.D., Purves, D., 1997. Correlated size variations in human visual cortex, lateral geniculate nucleus, and optic tract. J. Neurosci. 17, 2859–68.

Arcaro, M.J., Pinsk, M.A., Kastner, S., 2015. The Anatomical and Functional Organization of the Human Visual Pulvinar. J. Neurosci. 35, 9848–9871. doi:10.1523/JNEUROSCI.1575-14.2015

Arnold, P., Hill, F., 2001. Bisensory augmentation: A speechreading advantage when speech is clearly audible and intact. Br. J. Psychol. 92, 339–355. doi:10.1348/000712601162220

Aschenberner, B., Weiss, C., 2005. Phoneme-Viseme Mapping for German Video-Realistic Audio-Visual-Speech-Synthesis, in: Institut Für Kommunikationsforschung Und Phonetik, Universität Bonn.

Ashburner, J., 2007. A fast diffeomorphic image registration algorithm. Neuroimage 38, 95–113. doi:10.1016/J.NEUROIMAGE.2007.07.007

Azzopardi, P., Cowey, A., 1996. The overrepresentation of the fovea and adjacent retina in the striate cortex and dorsal lateral geniculate nucleus of the macaque monkey. Neuroscience 72, 627–639. doi:10.1016/0306-4522(95)00589-7

Bernstein, L.E., Tucker, P.E., Demorest, M.E., 2000. Speech perception without hearing. Percept. Psychophys. 62, 233–252. doi:10.3758/BF03205546

Blank, H., Anwander, A., von Kriegstein, K., 2011. Direct structural connections between voice- and face-recognition areas. J. Neurosci. 31, 12906–12915. doi:10.1523/JNEUROSCI.2091-11.2011

Blank, H., Davis, M.H., 2016. Prediction Errors but Not Sharpened Signals Simulate Multivoxel fMRI Patterns during Speech Perception. PLoS Biol. 14, e1002577. doi:10.1371/journal.pbio.1002577

Blank, H., von Kriegstein, K., 2013. Mechanisms of enhancing visual–speech recognition by prior auditory information. Neuroimage 65, 109–118. doi:10.1016/j.neuroimage.2012.09.047

Briggs, F., Usrey, W.M., 2008. Emerging views of corticothalamic function. Curr. Opin. Neurobiol. 18, 403–407. doi:10.1016/j.conb.2008.09.002

Briggs, F., Usrey, W.M., 2007. A fast, reciprocal pathway between the lateral geniculate nucleus and visual cortex in the macaque monkey. J. Neurosci. 27, 5431–5436. doi:10.1523/JNEUROSCI.1035-07.2007

Büchel, C., Turner, R., Friston, K., 1997. Lateral geniculate activations can be detected using intersubject averaging and fMRI. Magn. Reson. Med. 38, 691–694. doi:10.1002/mrm.1910380502

Camarillo, L., Luna, R., Nácher, V., Romo, R., 2012. Coding perceptual discrimination in

the somatosensory thalamus. Proc. Natl. Acad. Sci. U. S. A. 109, 21093–21098. doi:10.1073/pnas.1219636110

Campbell, R., Zihl, J., Massaro, D., Munhall, K., Cohen, M.M., 1997. Speechreading in the akinetopsic patient, L.M. Brain 120, 1793–1803. doi:10.1093/brain/120.10.1793

Chen, W., Zhu, X.H., Thulborn, K.R., Ugurbil, K., 1999. Retinotopic mapping of lateral geniculate nucleus in humans using functional magnetic resonance imaging. Proc Natl Acad Sci U S A 96, 2430–2434. doi:10.1073/pnas.96.5.2430

Christian, W., Esquembre, F., Barbato, L., 2011. Open Source Physics. Science 334, 1077–1078. doi:10.1126/science.1196984

Cudeiro, J., Sillito, A.M., 2006. Looking back: Corticothalamic feedback and early visual processing. Trends Neurosci. 29, 298–306. doi:10.1016/j.tins.2006.05.002

Cudeiro, J., Sillito, A.M., 1996. Spatial frequency tuning of orientation-discontinuity-sensitive corticofugal feedback to the cat lateral geniculate nucleus. J. Physiol. 490, 481–492. doi:10.1113/jphysiol.1996.sp021159

Díaz, B., Hintz, F., Kiebel, S.J., Von Kriegstein, K., 2012. Dysfunction of the auditory thalamus in developmental dyslexia. Proc. Natl. Acad. Sci. U. S. A. 109. doi:10.1073/pnas.1119828109

Eickhoff, S.B., Stephan, K.E., Mohlberg, H., Grefkes, C., Fink, G.R., Amunts, K., Zilles, K., 2005. A new SPM toolbox for combining probabilistic cytoarchitectonic maps and functional imaging data. Neuroimage 25, 1325–1335. doi:10.1016/j.neuroimage.2004.12.034

Foster, K.H., Gaska, J.P., Nagler, M., Pollen, D. a, 1985. Spatial and temporal frequency selectivity of neurones in visual cortical areas V1 and V2 of the macaque monkey. J. Physiol. 365, 331–363. doi:10.1113/jphysiol.1985.sp015776

Friederici, A.D., Alter, K., 2004. Lateralization of auditory language functions: a dynamic dual pathway model. Brain Lang. 89, 267–276. doi:https://doi.org/10.1016/S0093-934X(03)00351-1

Friston, K., 2005. A theory of cortical responses. Philos. Trans. R. Soc. B Biol. Sci. 360, 815–836. doi:10.1098/rstb.2005.1622

Friston, K.J., Ashburner, J., Kiebel, S.J., Nichols, T.E., Penny, W.D., 2007. Statistical parametric mapping: The analysis of functional brain images. Academic Press/Elsevier, Amsterdam.

Friston, K.J., Josephs, O., Zarahn, E., Holmes, A.P., Rouquette, S., Poline, J.-B., 2000. To Smooth or Not to Smooth? Bias and Efficiency in fMRI Time-Series Analysis. Neuroimage 12, 196–208. doi:10.1006/nimg.2000.0609

Friston, K.J., Zarahn, E., Josephs, O., Henson, R.N., Dale, A.M., 1999. Stochastic designs in event-related fMRI. Neuroimage 10, 607–619. doi:10.1006/nimg.1999.0498

Ghazanfar, A.A., Nicolelis, M.A., 2001. The structure and function of dynamic cortical and thalamic receptive fields. Cereb. Cortex 11, 183–193. doi:10.1093/cercor/11.3.183

Ghodrati, M., Khaligh-Razavi, S.-M., Lehky, S.R., 2017. Towards building a more complex view of the lateral geniculate nucleus: recent advances in understanding its role. Prog. Neurobiol. doi:10.1016/j.pneurobio.2017.06.002

Giraud, A.-L., Lorenzi, C., Ashburner, J., Wable, J., Johnsrude, I., Frackowiak, R.,

Kleinschmidt, A., 2000. Representation of the Temporal Envelope of Sounds in the Human Brain. J. Neurophysiol. 84, 1588–1598. doi:10.1093/cercor/11.3.183

Giraud, A.L., Price, C.J., Graham, J.M., Truy, E., Frackowiak, R.S.J., 2001. Cross-modal plasticity underpins language recovery after cochlear implantation. Neuron 30, 657–663. doi:10.1016/S0896-6273(01)00318-X

Guillery, R.W., Sherman, S.M., 2002. Thalamic Relay Functions and Their Role in Corticocortical Communication: Generalizations from the Visual System. Neuron 33, 163–175. doi:10.1016/S0896-6273(01)00582-7

Hickok, G., Poeppel, D., 2007. The cortical organization of speech processing. Nat. Rev. Neurosci. 8, 393–402. doi:doi:10.1038/nrn2113

Hicks, T.P., Lee, B.B., Vidyasagar, T.R., 1983. The responses of cells in macaque lateral geniculate nucleus to sinusoidal gratings. J. Physiol. 337, 183–200. doi:10.1113/jphysiol.1983.sp014619

Jesse, A., Massaro, D.W., 2010. The temporal distribution of information in audiovisual spoken-word identification. Atten. Percept. Psychophys. 72, 209–225. doi:10.3758/APP.72.1.209

Jones, E.G., 1985. The thalamus. Plenum Press, New York.

Kastner, S., O'Connor, D.H., Fukui, M.M., Fehd, H.M., Herwig, U., Pinsk, M.A., 2004. Functional imaging of the human lateral geniculate nucleus and pulvinar. J. Neurophysiol. 91, 438–448. doi:10.1152/jn.00553.2003

Kiebel, S.J., Daunizeau, J., Friston, K.J., 2008. A hierarchy of time-scales and the brain. PLoS Comput. Biol. 4, e1000209. doi:10.1371/journal.pcbi.1000209

Krupa, D.J., Ghazanfar, A.A., Nicolelis, M.A., 1999. Immediate thalamic sensory plasticity depends on corticothalamic feedback. Proc. Natl. Acad. Sci. U. S. A. 96, 8200–8205. doi:10.1073/pnas.96.14.8200

Lal, R., Friedlander, M.J., 1989. Gating of retinal transmission by afferent eye position and movement signals. Science 243, 93–96. doi:10.1126/science.2911723

Lee, S., Carvell, G.E., Simons, D.J., 2008. Motor modulation of afferent somatosensory circuits. Nat. Neurosci. 11, 1430–1438. doi:doi:10.1038/nn.2227

Ling, S., Pratte, M.S., Tong, F., 2015. Attention alters orientation processing in the human lateral geniculate nucleus. Nat. Neurosci. 18, 496–498. doi:10.1038/nn.3967

MacLeod, A., Summerfield, Q., 1987. Quantifying the contribution of vision to speech perception in noise. Br. J. Audiol. 21, 131–141. doi:10.3109/03005368709077786

Makinson, C.D., Huguenard, J.R., 2015. Attentional flexibility in the thalamus: now we're getting SOMwhere. Nat. Neurosci. 18, 2–4. doi:10.1038/nn.3902

Malpeli, J.G., Baker, F.H., 1975. The representation of the visual field in the lateral geniculate nucleus of Macaca mulatta. J. Comp. Neurol. 161, 569–594. doi:10.1002/cne.901610407

Mcgurk, H., Macdonald, J., 1976. Hearing Lips and Seeing Voices. Nature 264, 746–748. doi:10.1038/264746a0

Mikl, M., Mareček, R., Hluštík, P., Pavlicová, M., Drastich, A., Chlebus, P., Brázdil, M., Krupa, P., 2008. Effects of spatial smoothing on fMRI group inferences. Magn. Reson. Imaging 26, 490–503. doi:10.1016/j.mri.2007.08.006

Moerel, M., De Martino, F., Uğurbil, K., Yacoub, E., Formisano, E., 2015. Processing of frequency and location in human subcortical auditory structures. Sci. Rep. 5, 17048. doi:10.1038/srep17048

Mullen, K.T., Thompson, B., Hess, R.F., 2010. Responses of the human visual cortex and LGN to achromatic and chromatic temporal modulations: an fMRI study. J. Vis. 10, 1–19. doi:10.1167/10.13.13

Murphy, P.C., Sillito, A.M., 1987. Corticofugal feedback influences the generation of length tuning in the visual pathway. Nature 329, 727–729. doi:10.1038/329727a0

Navarra, J., Soto-Faraco, S., 2007. Hearing lips in a second language: visual articulatory information enables the perception of second language sounds. Psychol. Res. 71, 4–12. doi:10.1007/s00426-005-0031-5

Noesselt, T., Tyll, S., Boehler, C.N., Budinger, E., Heinze, H.-J., Driver, J., 2010. Sound-induced enhancement of low-intensity vision: multisensory influences on human sensory-specific cortices and thalamic bodies relate to perceptual enhancement of visual detection sensitivity. J. Neurosci. 30, 13609–13623. doi:10.1523/JNEUROSCI.4524-09.2010

O'Connor, D.H., Fukui, M.M., Pinsk, M.A., Kastner, S., 2002. Attention modulates responses in the human lateral geniculate nucleus. Nat. Neurosci. 5, 1203–1209. doi:10.1038/nn957

Oldfield, R.C., 1971. The assessment and analysis of handedness: The Edinburgh inventory. Neuropsychologia 9, 97–113. doi:10.1016/0028-3932(71)90067-4

Poeppel, D., Emmorey, K., Hickok, G., Pylkkanen, L., 2012. Towards a New Neurobiology of Language. J. Neurosci. 32, 14125–14131. doi:10.1523/JNEUROSCI.3244-12.2012

Ross, L.A., Saint-Amour, D., Leavitt, V.M., Javitt, D.C., Foxe, J.J., 2007. Do you see what I am saying? Exploring visual enhancement of speech comprehension in noisy environment. Cereb. Cortex 17, 1147–1153. doi:10.1093/cercor/bhl024

Rouger, J., Lagleyre, S., Fraysse, B., Deneve, S., Deguine, O., Barone, P., 2007. Evidence that cochlear-implanted deaf patients are better multisensory integrators. Proc. Natl. Acad. Sci. U. S. A. 104, 7295–300. doi:10.1073/pnas.0609419104

Rouiller, E.M., Rodrigues-Dagaeff, C., Simm, G., De Ribaupierre, Y., Villa, A., De Ribaupierre, F., 1989. Functional organization of the medial division of the medial geniculate body of the cat: Tonotopic organization, spatial distribution of response properties and cortical connections. Hear. Res. 39, 127–142. doi:10.1016/0378-5955(89)90086-5

Saalmann, Y.B., Kastner, S., 2011. Cognitive and Perceptual Functions of the Visual Thalamus. Neuron 71, 209–223. doi:10.1016/j.neuron.2011.06.027

Schneider, K.A., 2011. Subcortical mechanisms of feature-based attention. J. Neurosci. 31, 8643–8653. doi:10.1523/JNEUROSCI.6274-10.2011

Schneider, K.A., Kastner, S., 2009. Effects of sustained spatial attention in the human lateral geniculate nucleus and superior colliculus. J. Neurosci. 29, 1784–1795. doi:10.1523/JNEUROSCI.4452-08.2009

Schneider, K.A., Richter, M.C., Kastner, S., 2004. Retinotopic organization and functional subdivisions of the human lateral geniculate nucleus: a high-resolution functional magnetic resonance imaging study. J. Neurosci. 24, 8975–8985.

doi:10.1523/JNEUROSCI.2413-04.2004

Shannon, R. V, Zeng, F.-G.G., Kamath, V., Wygonski, J., Ekelid, M., 1995. Speech recognition with primarily temporal cues. Science 270, 303–4. doi:10.1126/science.270.5234.303

Sillito, A.M., Cudeiro, J., Murphy, P.C., 1993. Orientation sensitive elements in the corticofugal influence on centre-surround interactions in the dorsal lateral geniculate nucleus. Exp. Brain Res. 93, 6–16. doi:10.1007/BF00227775

Sillito, A.M., Jones, H.E., Gerstein, G.L., West, D.C., 1994. Feature-linked synchronization of thalamic relay cell firing induced by feedback from the visual cortex. Nature 369, 479–482. doi:10.1038/369479a0

Studebaker G.A., 1985. A "rationalized" arcsine transform. J. Speech Hear. Res. 28, 455–462. doi:10.1044/jshr.2803.455

Suga, N., Ma, X., 2003. Multiparametric corticofugal modulation and plasticity in the auditory system. Nat. Rev. Neurosci. 4, 783–794. doi:10.1038/nrn1222

Sumby, W.H., Pollack, I., 1954. Visual Contribution to Speech Intelligibility in Noise. J. Acoust. Soc. Am. 26, 212–215. doi:10.1121/1.1907309

Sylvester, R., Haynes, J.D., Rees, G., 2005. Saccades differentially modulate human LGN and V1 responses in the presence and absence of visual stimulation. Curr. Biol. 15, 37–41. doi:10.1016/j.cub.2004.12.061

Temereanca, S., Simons, D.J., 2004. Functional topography of corticothalamic feedback enhances thalamic spatial response tuning in the somatosensory whisker/barrel system. Neuron 41, 639–651. doi:10.1016/S0896-6273(04)00046-7

Thompson, S.K., von Kriegstein, K., Deane-Pratt, A., Marquardt, T., Deichmann, R., Griffiths, T.D., McAlpine, D., 2006. Representation of interaural time delay in the human auditory midbrain. Nat. Neurosci. 9, 1096–1098. doi:doi:10.1038/nn1755

van Wassenhove, V., Grant, K.W., Poeppel, D., 2005. Visual speech speeds up the neural processing of auditory speech. Proc. Natl. Acad. Sci. U. S. A. 102, 1181–1186. doi:10.1073/pnas.0408949102

von Kriegstein, K., Patterson, R.D., Griffiths, T.D., 2008. Task-dependent modulation of medial geniculate body is behaviorally relevant for speech recognition. Curr. Biol. 18, 1855–1859. doi:10.1016/j.cub.2008.10.052

Warren, P., Marslen-Wilson, W., 1987. Continuous uptake of acoustic cues in spoken word recognition. Percept. Psychophys. 41, 262–275. doi:10.3758/BF03208224

Winner, J.A., 1984. The human medial geniculate body. Hear. Res. 15, 225–247. doi:10.1016/0378-5955(84)90031-5

Wunderlich, K., Schneider, K.A., Kastner, S., 2005. Neural correlates of binocular rivalry in the human lateral geniculate nucleus. Nat. Neurosci. 8, 1595–1602. doi:10.1038/nn1554

Zhang, Y., Suga, N., Yan, J., 1997. Corticofugal modulation of frequency processing in bat auditory system. Nature 387, 900–903. doi:10.1038/43180

# Supplementary Information

## Supplementary Methods

### Training Experiment 1

To enable participants to perform the identity tasks during fMRI scanning in Experiment 1, they were trained on the audio-visual identity of the three speakers and the three cell phones, directly before entering the MRI-scanner. The stimulus material used for the training was different from the material used for the fMRI experiment but training and experimental stimuli were acquired from the same speakers (three native German male speakers) and cell phones (three different brands) during the same recording session and with the same apparatus and settings. In the person videos, each speaker said eight two-syllable words and twelve semantically neutral and syntactically homogeneous five-word sentences (Example: "Der Junge trägt einen Koffer", English: "The boy carries a suitcase"). Keypad tone samples of each cell phone included eight sequences of two to five key presses and twelve sequences of six to nine sequences.

For the training, participants first viewed audio-visual videos of the speakers (learning phase) and the cell phones, and then were tested in their knowledge of the speakers and cell phone identities (test phase). For the learning phase in the person condition, there were twelve audiovisual videos of each speaker and each video contained a different five-word sentence (mean duration 3.10 s ±0.40). The sentences were the same for all speakers. Videos were randomly presented and the name of each speaker was written on the screen during the presentation of the video. Between the videos a fixation cross appeared on the screen for 1 s. Participants were instructed to learn the association between three speaker's names, faces, and voices. Participants learned the cell phone identities in a similar manner to the person identities. Twelve audiovisual videos for each cell phone were randomly presented which differed in the number of key presses, from six to nine, and the specific keys pressed (mean duration 3.31 s ±1.06). An asterisk was presented on the screen for 1 s between the videos. Participants were asked to learn the association between the cell-phone image and keypad tones. Participants were explicitly told before the learning phase that they would be tested on their knowledge of the name-voice-face and cell phone-identity-key tone associations after the learning phase.

In the recognition test phase, visual-only and auditory-only stimuli were created from the audiovisual videos. The test phase for the person identities included 8 words visual-only (mean duration 1.74 s ±0.16) and auditory-only (mean duration 0.89 ±0.11). The test phase for the cell phone identities included 8 key presses sequences visual-only (mean duration 1.64 s ±0.23) and auditory-only (mean duration 1.66 s ±0.24). Participants first watched a muted video of a person saying a two-syllable word (or a sequence of two to five key presses on a cell phone keypad) and subsequently listened to a voice saying the given word (or a key-tone sequence). Between the visual and auditory stimuli an asterisk was present for 1 s. Participants were asked to indicate whether the auditory voice (or key-tone) belonged to the face (or cell phone) in the previous video. Participants received feedback on responses that were correct, incorrect, and too slow (for responses > 2 s after onset of the second stimulus). The feedback was presented for 600 ms. Trials were separated by an interstimulus interval of 2 s. The training, including the learning and the test, took ca. 25 minutes. Training was repeated twice for all participants (50 minutes in total). If a participant performed less than 80% correct after the second training session, the training was repeated a third time.

*Eye tracking data analysis*

Normal distribution of the data was assessed with a Shapiro-Wilk normality test. The analysis included the data from 15 (of the total 16) participants because for one participant there were difficulties with the calibration of the eye tracker. For another participant only the eye tracking data for the first run were available, because track of the eye was lost during the second run.

We first computed the average number and duration of fixations for each participant and each type of task (visual-speech recognition task and face-identity task). Fixations were defined as periods in which eye movement dispersion did not exceed 1° at least for 100 ms (Hannula and Ranganath, 2009; Wolf et al., 2014). Blocks in which the eye tracker lost fixation were excluded from analyses (10.46% in the visual-speech recognition task and 11.55% in the face-identity task). The average number and duration of fixations for both tasks followed a normal distribution (all Ws(15) > 0.92, all ps > 0.05) and eye measures for each task were compared by means of a paired samples t-test.

In a second step, we computed the average number and duration of fixations for each participant and each type of task within four regions of interest (ROI): the speakers' eyes, nose, mouth, and the rest of the video. This analysis served to assess whether there might be eye-movement differences between the two task conditions, depending on the ROI. The face ROIs (i.e., eyes, nose, and mouth) had similar sizes. Together they covered most of the face of the speakers (except for the tip of the chin and the top of the forehead) (Supplementary Fig. 1). We included in the analysis only those blocks for which at least 75% of the fixations fell within the face ROIs (i.e., the eyes, mouth, and nose) to control for drifts in eye movement position caused by the participants slow drift of head movement and the lack of recalibration of the eye-tracker in between sessions. Note that the average number of fixations should not be influenced by slow head movement since eye movements (i.e., end of fixations) were considered as fast movements of 1º within 100 ms. However, when analyzing eye position the coordinates are greatly influenced by slow drifts of the head that carry misallocations of the eyes from block to block. Eight participants were excluded from the analysis because only 1 block for one experimental task survived the 75% criterion of fixations within the face ROIs. For the remaining seven participants there was an average of 14.14 (±6.54) blocks for the visual-speech recognition task and 11.57 (±5.99) for the face-identity task that survived the 75% criterion of fixations within the face ROIs. The average number of fixations for each ROI and experimental task followed a normal-distribution (all $Ws(7) > 0.81$, all $ps > 0.05$). We, therefore, analyzed the eye position for each task by means of a repeated-measures ANOVA on the number of fixations with the factors ROI (eye, nose, and mouth) and experimental task (visual-speech recognition and face-identity tasks).

# Supplementary Figures

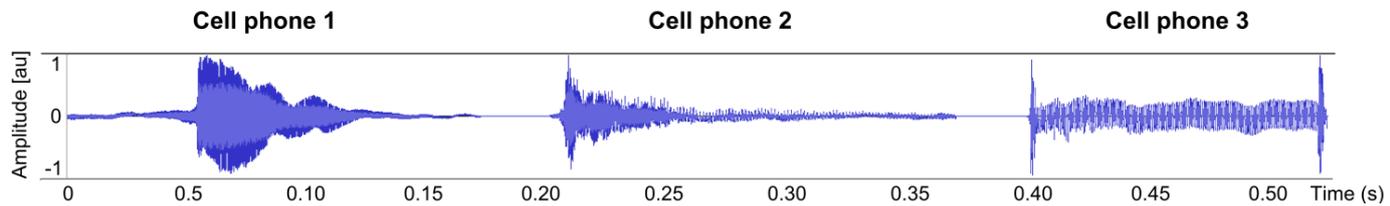

**Figure 1. Experiment 1: Cell phone key tones.** Waveform graphs of the key tones used as auditory stimuli for the cell phone conditions. Graphs were created with the software Audacity (2.0.3).

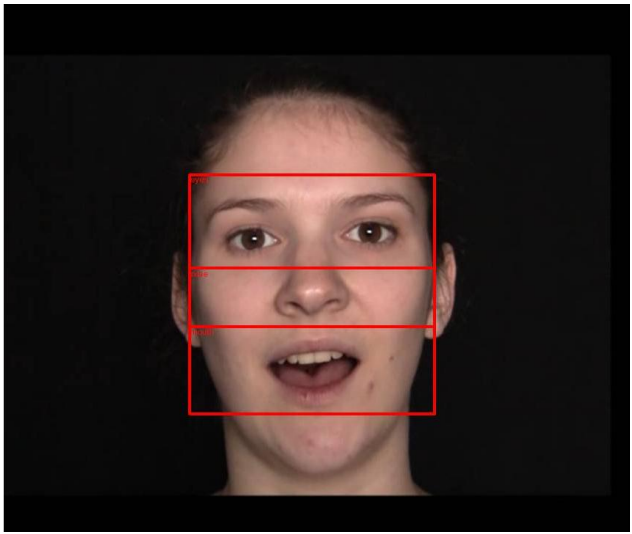

**Figure 2. Experiment 2: Regions of interest (ROIs) for the eye tracking data analysis.** The figure displays the eye, nose, and mouth ROIs for the eye tracking data analysis overlaid on an image extracted from one of the videos used as stimuli for Experiment 2.

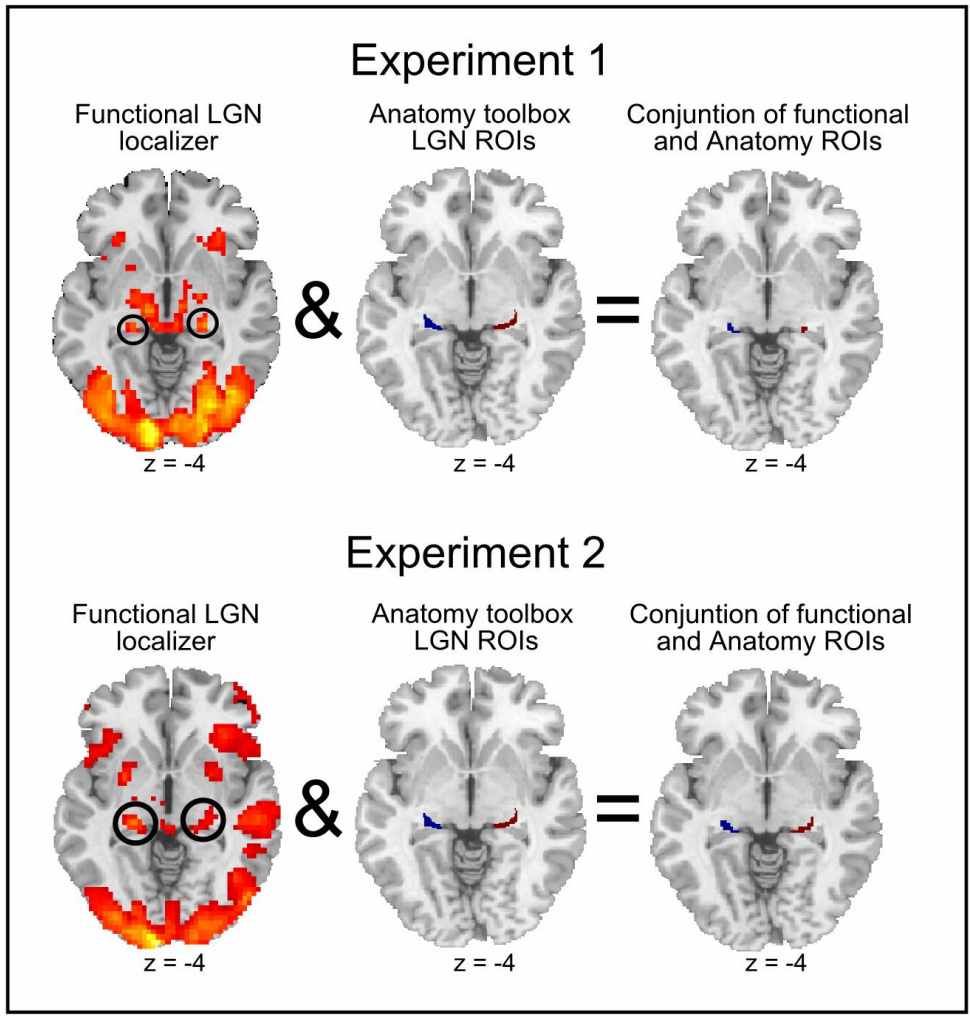

**Figure 3. Definition of the regions of interest.** We functionally localized the LGNs at the group level by means of the independent functional localizers. The clusters localized in the anatomical position of the LGNs were then combined with probabilistic cytoarchitectonic maps of the LGNs. The intersections of the group functional clusters and the probabilistic maps were used as regions of interest for the analyses of the experimental conditions.

## Supplementary Tables

**Table 1. LGN coordinates reported by fMRI studies.**

| Study<br>Experiment | MNI Local maxima (x,y,z) | |
|---|---|---|
| | Right LGN | Left LGN |
| _Present study_ | | |
| Experiment 1: localizer, n=18 | 24, -27, -3 | -21, -27, -3 |
| Experiment 1: task modulation, n=18 | 24, -27, -6 | -21, -27, -6 |
| Experiment 2: localizer, n=16 | 27, -24, -3 | -21, -27, -3 |
| Experiment 2: task modulation, n=16 | 24, -27, -3 | ----- |
| _O'connor et al. (2002)*, n = 4_ | | |
| Anatomical experiment | 24,-21,-11 | -24,-21,-10 |
| Enhancement experiment | 25,-19,-9 | -22,-31,-10 |
| Suppression experiment | 22,-20,-9 | -25,-21,-10 |
| Baseline experiment | 23,-20,-10 | -23,-19,-10 |
| _Kastner et al. (2004)*_ | | |
| Anatomical experiment, n=9 | 24,-20,-11 | -24,-20,-11 |
| Localizer, n=6 | 22,-18,-11 | -23,-20,-10 |
| Contrast, n=5 | 22,-23,-10 | -21,-20,-10 |
| Frequency, n=4 | 24,-22,-10 | -21,-22,-11 |
| _Schneider et al. (2004)*_ | | |
| Polar angle maps, n=7 | 24,-22,-9 | -22,-22,-9 |
| Eccentricity maps, n=7 | Not reported | Not reported |
| Contrast sensitivity maps, n=5 | Not reported | Not reported |
| _Wunderlich et al. (2005), n=5_ | Not reported | Not reported |
| _Schneider & Kastner (2009), n=4_ | Not reported | Not reported |
| _Mullen et al. (2010)*, n=6_ | 22,-25,-8 | -22,-24,-9 |
| _Schneider (2011), n=11_ | Not reported | Not reported |
| _Arcaro et al. (2015), n=9_ | 24,-25,-6 | -24,-25,-6 |

The table lists the average coordinates for the LGNs of the present study and those reported by previous studies. The Talairach coordinates reported by some studies (marked with an asterisk in the table) were transformed to MNI coordinates by means of the Yale BioImage Suite Package (http://sprout022.sprout.yale.edu/mni2tal/mni2tal.html). The MNI to Talairach conversion is based on Lacadie et al. (2008).

**Table 2. Comparison of the maximum statistics of the significant effects across different smoothing kernel sizes for the fMRI data.**

|  | Smoothing kernel size | Left LGN (MNI coordinates: x,y,z) | Right LGN (MNI coordinates: x,y,z) |
|---|---|---|---|
| **Experiment 1** |  |  |  |
| *Categorical Analysis 1* | 4 mm | -21, -27, -6 | 24, -27, -6 |
|  | 2 mm | -21, -27, -6<br>0 mm displacement | 24, -27, -6<br>0 mm displacement |
|  | 0 mm | -24, -27, -6<br>3 mm more lateral | 24, -27, -6<br>0 mm displacement |
| *Correlation Analysis 2* | 4 mm | -21, -27, -6 | 21, -27, -3 |
|  | 2 mm | -21, -30, -3<br>3 mm more anterior and 3 more superior | 24, -27, -6<br>3 mm more lateral and 3 more inferior |
|  | 0 mm | -21, -30, -3<br>3 mm more anterior and 3 more superior | 24, -27, -3<br>3 mm more lateral |
| **Experiment 2** |  |  |  |
| *Categorical Analysis 1* | 4 mm | n.s. | 24, -27, -3 |
|  | 2 mm | ----- | 24, -27, -3<br>0 mm displacement |
|  | 0 mm | ----- | 27, -27, -3<br>3 mm more lateral |

The table lists the maximum statistic for the contrast visual-speech recognition vs. face-identity task (Analysis 1) of Experiments 1 and 2. The results with 4 mm smoothing are reported in the main text. For the 2 mm and 0 mm smoothing, we report direction in the x, y, and z axis of the displacement and the amount (in mm) in relation to the coordinate reported in the manuscript for 4 mm smoothing.

## Supplementary references

Hannula, D.E., Ranganath, C., 2009. The Eyes Have It: Hippocampal Activity Predicts Expression of Memory in Eye Movements. Neuron 63, 592–599. doi:10.1016/j.neuron.2009.08.025

Lacadie, C.M., Fulbright, R.K., Rajeevan, N., Constable, R.T., Papademetris, X., 2008. More accurate Talairach coordinates for neuroimaging using non-linear registration. Neuroimage 42, 717–725. doi:10.1016/j.neuroimage.2008.04.240

Wolf, R.C., Philippi, C.L., Motzkin, J.C., Baskaya, M.K., Koenigs, M., 2014. Ventromedial prefrontal cortex mediates visual attention during facial emotion recognition. Brain 137, 1772–80. doi:10.1093/brain/awu063